\documentclass[aps,prd,twocolumn,
	superscriptaddress,
	showpacs,preprintnumbers,amsmath,amssymb,nofootinbib,
	longbibliography
	]{revtex4-1}
\usepackage[
	colorlinks=true,
	pdfstartview=FitV,
	linkcolor=blue,
	citecolor=magenta,
	urlcolor=blue,
	bookmarks=true,
	bookmarksnumbered=true,
	pdftitle={},
	pdfauthor={}
	]{hyperref}
\usepackage[dvips]{graphicx}
\usepackage{grffile}
\usepackage{braket}
\usepackage{bm,latexsym,amsmath,amssymb,amsfonts,mathrsfs,textcomp}
\usepackage[normalem]{ulem}  % \sout{old text} for strikeout
\usepackage{color}
\input{colordvi.tex}
\usepackage{tikz}
\usepackage{tikz-feynhand}

\newcommand{\be}{\begin{equation}}
\newcommand{\ee}{\end{equation}}
\newcommand{\bear}{\begin{eqnarray}}
\newcommand{\eear}{\end{eqnarray}}
\newcommand{\ba}{\begin{array}}
\newcommand{\ea}{\end{array}}

\newcommand{\diff}{\mathrm{d}} 
\newcommand{\rmi}{\mathrm{i}} 
\newcommand{\rme}{\mathrm{e}} 
\newcommand{\bp}{\bm{p}} 
\newcommand{\bk}{\bm{k}} 
\newcommand{\bnab}{\bm{\nabla}}

\newcommand{\im}{\mathop{\mathrm{Im}}}
\newcommand{\re}{\mathop{\mathrm{Re}}}
\newcommand{\tilgamma}{\widetilde{\gamma}}

\newcommand\Lcal{\mathcal{L}}

\newcommand\Mcal{\mathcal{M}}
\newcommand\pu{\bar{u}}
\newcommand\pLam{\bar{\Lambda}}
\newcommand\pSig{\bar{\Sigma}}

\newcommand{\cout}[1]{ \if 0 {#1} \fi }

\def\({\left(}
\def\){\right)}

\newcommand{\matrixT}{\mathcal T}

\begin{document}
\preprint{KEK-TH-2565, J-PARC-TH-0294}

\author{Yoshimasa Hidaka}
\email{hidaka@post.kek.jp}
\affiliation{KEK Theory Center, Tsukuba 305-0801, Japan}
\affiliation{Graduate University for Advanced Studies (Sokendai), Tsukuba 305-0801, Japan}
\affiliation{Department of Physics, The University of Tokyo, 7-3-1 Hongo, Bunkyo-ku, Tokyo  113-0033, Japan}
\affiliation{International Center for Quantum-field Measurement Systems for Studies of the Universe and Particles (QUP), KEK, Tsukuba, 305-0801, Japan}
\affiliation{RIKEN iTHEMS, RIKEN, Wako 351-0198, Japan}

\author{Masaru Hongo}
\email{hongo@phys.sc.niigata-u.ac.jp}
\affiliation{Department of Physics, Niigata University, Niigata 950-2181, Japan}
\affiliation{RIKEN iTHEMS, RIKEN, Wako 351-0198, Japan}

\author{Mikhail A. Stephanov}
\email{misha@uic.edu}
\affiliation{Department of Physics, University of Illinois, Chicago, Illinois 60607,  USA}
\affiliation{Kadanoff  Center  for  Theoretical  Physics,  University  of  Chicago,  Chicago,  Illinois  60637,  USA}
\affiliation{Laboratory for Quantum Theory at the Extremes, University of Illinois, Chicago, Illinois 60607,  USA}

\author{Ho-Ung Yee}
\email{hyee@uic.edu}
\affiliation{Department of Physics, University of Illinois, Chicago, Illinois 60607,  USA}
\affiliation{Laboratory for Quantum Theory at the Extremes, University of Illinois, Chicago, Illinois 60607,  USA}

\date{\today}
\title{
Spin relaxation rate for baryons in thermal pion gas
}

\begin{abstract}
We study the relaxation dynamics of the spin polarization of baryons (nucleon and $\Lambda$-baryon), in a thermal pion gas as a simple model of the hadronic phase of the QCD plasma produced in relativistic heavy-ion collisions.
For this purpose, we formulate the quantum kinetic theory for the spin density matrix of baryons in the leading order of the gradient expansion.
Considering the baryon-pion elastic scattering processes as the dominant interaction between baryons and thermal pions, we compute the spin relaxation rate of nucleons and $\Lambda$-baryons in a pion gas up to temperature 200 MeV.
In the case of nucleons, we evaluate the spin relaxation rate in the $s$-channel resonance approximation, based on the known experimental data on $\Delta$-resonances.
We also estimate the spin relaxation rate for $\Lambda$-baryons, based on experimental inputs and theoretical models for the low-energy $\Lambda$$\pi$ scattering, including the chiral perturbation theory.
\end{abstract}
\maketitle

% \tableofcontents

\section{Introduction \label{sec1}}

Experimentally observed spin polarization of hadrons aligned with the orbital angular momentum of the colliding nuclei ~\cite{STAR:2017ckg,STAR:2019erd,ALICE:2019aid,STAR:2020xbm} is the manifestation of a quantum mechanical response to the strong vorticity of the QCD plasma created in relativistic heavy-ion collisions. 
Broadly, theoretical predictions based on local thermal equilibrium of spin polarization and fluid vorticity are in reasonable agreement with the experimental observation.
However, any quantitative comparison in detail clearly requires a more sophisticated treatment, which should take into account the time evolution of spin polarization locally out-of-equilibrium.
Currently, two mainstream approaches for addressing the potentially out-of-equilibrium dynamics of spin polarization are under intense investigation: kinetic theory incorporating spin~\cite{Gao:2019znl,Weickgenannt:2019dks,Hattori:2019ahi,Wang:2019moi,Li:2019qkf,Kapusta:2019sad,Liu:2019krs,Yang:2020hri,Liu:2020flb,Weickgenannt:2020aaf,Weickgenannt:2021cuo,Sheng:2021kfc,Lin:2021mvw,Hu:2021pwh,Sheng:2022ssd,Weickgenannt:2022zxs,Fang:2022ttm,Wang:2022yli,Dong:2022yzt,Wagner:2022amr} (see also Ref.~\cite{Hidaka:2022dmn} for a recent review on the quantum kinetic theory), and hydrodynamics incorporating spin~\cite{Florkowski:2017ruc,Florkowski:2018fap,Hattori:2019lfp,Fukushima:2020ucl,Bhadury:2020puc,Shi:2020htn,Li:2020eon,Gallegos:2021bzp,Liu:2021uhn,She:2021lhe,Hongo:2021ona,Peng:2021ago,Gallegos:2022jow,Cao:2022aku,Weickgenannt:2022qvh,Biswas:2023qsw}.

Obviously, the key quantity in such out-of-equilibrium time-evolution is the relaxation rate of spin polarization toward its equilibrium value.
The spin polarization in general is not a conserved quantity, and thus, gives one of the non-hydrodynamic modes not amenable to a hydrodynamics description.
In the specific case where spin polarization shows much slower relaxation than other non-hydrodynamic modes, a quasi-hydrodynamics or Hydro+ description~\cite{Stephanov:2017ghc}, dubbed as the spin hydrodynamics, can be introduced to include the spin polarization as an additional quasi-hydrodynamic variable.
The spin hydrodynamics, in this case, is characterized by a new transport coefficient, called rotational viscosity, which corresponds to the spin relaxation rate $\gamma_S$.
In Ref.~\cite{Hongo:2022izs}, the spin relaxation rate for heavy quarks in a high-temperature weakly coupled quark-gluon plasma was computed to the leading-order of QCD coupling constant $\alpha_s$. 
It behaves as $\gamma_S\sim \alpha_s^2\log(1/\alpha_s) \times({T^3/ M_q^2})$ as a function of the heavy-quark mass $M_q(\gg T)$ and the temperature $T$.

In this work, we extend our study of spin relaxation in QCD plasma to a low-temperature hadronic phase, focusing on the spin polarization of baryons, e.g., nucleons and $\Lambda$ baryons.
It is directly motivated by the experimental observation of the spin-polarized $\Lambda$-baryons in RHIC~\cite{STAR:2017ckg,STAR:2019erd,ALICE:2019aid,STAR:2020xbm}.
The QCD plasma in the hadronic phase at low temperature and vanishing baryon chemical potential is mostly a gas of weakly interacting pions, described by chiral perturbation theory~\cite{Weinberg:1978kz,Gasser:1983yg,Gasser:1984gg} (see, e.g., Ref.~\cite{Scherer-Schindler2011} for a recent pedagogical review). 
The chiral perturbation theory, however, becomes inadequate around $T \simeq 100$~MeV, although pions are still the major component of the plasma. Here $f$ represents the pion decay constant.
As a simple model for the hadronic phase in this work, we will therefore approximate the plasma by a thermal pion gas without heavy use of the chiral perturbation theory.
We note that our approach works well for a low-temperature phase, while it eventually breaks down as the temperature becomes closer to the crossover temperature $T_c\sim 200$ MeV from the hadronic to quark-gluon plasma phases.

The baryon density in this temperature range is dilute, and thus, we will neglect inter-baryon interactions in the plasma.
In other words, we treat each baryon independent of other constituents of the plasma, except thermal pions surrounding it.
The dominant baryon-pion interaction in this regime is the elastic two-body scattering of the baryon with background pions~\cite{Leutwyler:1990uq}. 
This motivates us to formulate the quantum kinetic theory of baryons interacting with the surrounding pions.
Based on this picture, we will construct the leading-order collision operator in the quantum kinetic theory of the spin density matrix, which allows us to compute the spin relaxation rate from the spin-flipping scattering processes.

Let us briefly summarize our picture of the relevant degrees of freedom and processes in the following.
At a sufficiently low temperature, the baryon-pion scattering amplitude is describable in the chiral perturbation theory. 
Indeed, there exists a well-developed chiral perturbation theory including nucleons (see, e.g., Ref.~\cite{Scherer-Schindler2011} for a pedagogical review). 
Above $T=100$ MeV, this is no longer a good description of the nucleon-pion scattering amplitudes because the  $\Delta$-baryon resonance starts to dominate the scattering amplitudes.
The nucleon-pion scattering amplitudes in this energy range have been experimentally measured with a good precision a while ago---indeed, this was how the $\Delta$-baryon resonance was discovered. 
Thus, instead of relying on the chiral perturbation theory, we can directly utilize these experimental data to evaluate the spin-flipping amplitudes that are needed to compute the spin relaxation rate.

On the other hand, experimental data on the $\Lambda\pi$ scattering amplitude is absent.
We will thus resort to a reasonable modeling, taking into account the known hadron spectrum including nearby strange baryons as well as the $s$-channel resonances in the $\Lambda\pi$ scattering. 
Specifically, we consider two most relevant strange baryons, i.e., $\Sigma$-baryon (of average mass 1190 MeV) and $\Sigma^*(1385)$. 
The former should be considered as a bound state of $\Lambda$-$\pi$ system in the total angular momentum $J=1/2$ channel, and the latter a resonance of a finite width in the $J=3/2$ channel. 
Fortunately for us, the width of $\Sigma^*(1385)$ is known experimentally from its dominant decay to $\Lambda+\pi$, i.e., $\Gamma_{\Sigma^*}=36$ MeV, which can be used to evaluate the spin relaxation rate for the $\Lambda$-baryon.
Based on these inputs, we are able to compute the spin relaxation rates of nucleons and $\Lambda$-baryons numerically, up to temperature $T=200$ MeV.

The organization of the paper is as follows.
In Sec.~\ref{sec:spin-kinetic-theory}, we formulate the quantum kinetic theory with the spin density matrix to describe the baryon spin relaxation. 
We also derive the formulae for the spin relaxation rate of nucleons and $\Lambda$-baryon in terms of the scattering amplitude.
In Sec.~\ref{sec:spin-relaxation}, we evaluate those spin relaxation rates mainly relying on the experimental data.
Section~\ref{sec:Discussion} is devoted to the discussion.
In Appendix \ref{sec:Kadanoff-Baym}, we give an alternative derivation of the spin kinetic theory of baryons based on the Kadanoff-Baym formalism.
Throughout the present paper, we use the unit system where $\hbar=1$ and $c = 1$.

\section{Spin density matrix and its time evolution}
\label{sec:spin-kinetic-theory}

In this section, we provide a density-matrix formulation to describe the spin relaxation of baryons in the pion gas following the formulation developed in Ref.~\cite{Li:2019qkf}. 
After introducing the spin density matrix in Sec.~\ref{sec:density-matrix}, we apply it to derive the formulae for the spin relaxation rate of nucleons in terms of the scattering amplitude in Sec.~\ref{sec:nucleon} and $\Lambda$-baryons in Sec.~\ref{sec:Lambda}.
In Appendix \ref{sec:Kadanoff-Baym}, we present an alternative route based on the Kadanoff-Baym formalism including spin degrees of freedom \cite{Kadanoff-Baym1961}.

\subsection{Spin density matrix for baryons}
\label{sec:density-matrix}

Let us consider a baryon in the low-temperature pion gas.
In the limit where the baryon density is dilute, we can neglect baryon-baryon interactions compared to the leading-order baryon-pion interactions. 
In this case, we can introduce a notion of the density matrix operator $\hat\rho$ for one baryon defined in the Hilbert space of the single baryon state.
Expanding this baryon density operator in the basis of momentum and spin eigenstates $\{|\bm p,s\rangle\}$ with $s=\pm 1/2$, we can, in general, express $\hat\rho$ as
\be
\hat\rho
= \int_{\bm p_1,\bm p_2}
\sum_{s_1,s_2} |\bm p_1,s_1 \rangle 
\rho({\bm p_1},s_1;{\bm p_2},s_2) 
\langle{\bm p_2},s_2|,
\ee 
in terms of a function $\rho({\bm p_1},s_1;{\bm p_2},s_2)$.
Here, we introduced a shorthand notation for the momentum integral as 
\be
\int_{\bp_1, \cdots \bp_n}\equiv 
\int \frac{\diff^3 p_1}{(2\pi)^3} \cdots \frac{\diff^3 p_n}{(2\pi)^3}.
\ee
Equivalently, we can represent the same information in the mixed position-momentum basis by performing a Wigner transform as
\be
 \rho(\bm x,\bm p) 
 \equiv
  \int_{\bm p_r} \rho(\bm p+{\bm p}_r/2,\bm p-{\bm p}_r/2)
  \rme^{\rmi \bm p_r\cdot\bm x},
\ee 
where we have omitted the spin variables for simplicity.

Since we consider the single baryon dynamics in the approximately uniform pion gas, $\bm{x}$-dependence of the Wigner transformed density matrix $\rho(\bm x,\bm p)$ is assumed to be smooth enough.
Then, the baryon-pion interactions, which give rise to spin relaxation of baryons, happen at a length scale that is much shorter than the scale of variation in $\bm x$ of the density matrix.
As a result, we can neglect $\bm x$-dependence in $\rho(\bm x,\bm p)$ in the leading order of the gradient expansion. 
This amounts to assuming that the density matrix is approximately diagonal in the momentum space
\be
 \rho({\bm p_1},s_1;{\bm p_2},s_2)
 \simeq \rho(\bm p_1;s_1,s_2)
 (2\pi)^3 \delta^{(3)} (\bm p_1-\bm p_2).
 \label{eq:rho}
\ee

We here note that the $2\times 2$ spin density matrix 
$\rho(\bm p;s,s')\equiv [\rho^{2\times 2}(\bm p)]_{s,s'}$ 
serves as the distribution function in momentum space with matrix indices $s$ and $s'$. 
When it is summed over the two spin states, the result should be identified as the probability distribution function in the ordinary kinetic theory~\cite{Landau:Kinetic} as $n(\bm p)= \sum_s \rho(\bm p;s,s)=\mathrm{Tr} \left(\rho^{2\times 2}(\bm p)\right)$.
Similarly, one can obtain the spin distribution in momentum space by computing the expectation value of the spin operator given by $\bm S(\bm p)=({\hbar/ 2})\mathrm{Tr}\left(\bm\sigma\rho^{2\times 2}(\bm p)\right)$, where $\bm \sigma$ are the Pauli matrices.%
\footnote{
In this definition of the spin distribution we write $\hbar$ explicitly. 
We will, however, use the unit system with $\hbar=1$ in the reminder of the paper.
}
%Using these distribution functions, we can decompose the $2\times 2$ spin density matrix as 
%\begin{equation}
% \rho^{2\times 2}(\bm p)
%  = {\frac12}f(\bm p){\bf 1}_{2\times 2}
%  +\bm S(\bm p)\cdot\bm\sigma. 
%  \label{eq:spin-density-matrix}
%\end{equation}

Suppose that the baryon is put into a thermal bath of pion gas.
Neglecting other baryons and mesons, we introduce the total density operator at the initial time as 
$\hat{\rho}_{\mathrm{tot}} = \hat{\rho}\otimes\hat{\rho}_{\mathrm{eq}}^\pi$.
Here, $\hat{\rho}_{\mathrm{eq}}^\pi$ denotes the equilibrium density operator of the thermal pion gas.
The time evolution of the total density matrix is given by 
$\hat{\rho}_{\mathrm{tot}} (t+\Delta t) = \hat U(\Delta t) \hat{\rho}_{\mathrm{tot}} (t)\hat U^\dagger(\Delta t)$, where $\hat U(\Delta t)$ is the unitary time-evolution operator generated by the total Hamiltonian, including interactions between baryons and pions.

We now perform the average over the pions to obtain the effective time-evolution equation for the baryon density matrix $\hat\rho$.
The initial state for the pions is given by the thermal density matrix $\hat{\rho}_{\mathrm{eq}}^\pi$. 
We assume that the final states of the pions after interactions with the baryon do not matter for the time evolution of the baryon, due to fast thermal scrambling of the pions. 
Therefore, we take a sum over all possible final states of pions after interactions. 
Let us denote this thermal average and the summation over final states by a symbol 
$\langle \cdots \rangle_{\mathrm{eq}}^\pi$.

To see what the above procedure means, we now evaluate
\begin{equation}
 \mathrm{Tr}_{\pi} 
  \hat\rho_{\mathrm{tot}} (t+\Delta t)
  = \langle
  \hat U(\Delta t)
  \hat\rho (t)
  \hat U^\dagger(\Delta t)
  \rangle_{\mathrm{eq}}^\pi,
\end{equation}
after expanding $\hat U(\Delta t)$ and $\hat U^\dagger(\Delta t)$ in power series of the pion operators $\hat\pi(x)$.
Here, $\mathrm{Tr}_\pi$ denotes the trace over all possible final states of the pion in the background.
Then, we are led to compute the thermal average in the pion sector
\begin{align}
% &\langle 
%  \hat\pi_1(x_1)\hat\pi_1(x_2)\cdots
%  \hat\pi_2(y_1)\hat\pi_2(y_2)\cdots
% \rangle_{\mathrm{eq}}^\pi
% \nonumber \\
% =& 
 \mathrm{Tr}_{\pi}
 \left(
  \hat\pi(x_1)\hat\pi(x_2)\cdots
  \hat\rho_{\mathrm{eq}}^\pi
  \hat\pi(y_1)\hat\pi(y_2)\cdots
 \right) ,
\end{align}
where $\hat\pi(x_i)$ are the pion operator from $\hat U(\Delta t)$, and $\hat\pi(y_i)$ are those from $\hat U^\dagger(\Delta t)$.
One can find that this is equivalent to the correlation functions in the Schwinger-Keldysh contour~\cite{Schwinger:1960qe,Keldysh:1964ud} with an initial thermal density matrix $\hat\rho_{\mathrm{eq}}^\pi$ as
\begin{align}
 & \mathrm{Tr}_\pi
 \left( 
  \hat\pi(x_1)\hat\pi(x_2)\cdots
  \hat\rho_{\mathrm{eq}}^\pi \hat\pi(y_1)\hat\pi(y_2)
  \cdots
 \right)
 \nonumber \\
 =& \left\langle
  \pi_1(x_1)\pi_1(x_2)\cdots \pi_2(y_1)\pi_2(y_2)\cdots
  \right\rangle_\mathrm{SK},
\end{align}
where $\pi_{1,2}(x)$ are the pion fields placed in the forward or backward time contours, respectively.
In other words, we can consider the pion operators appearing in $\hat U(\Delta t)$ and $\hat U^\dagger(\Delta t)$ as the fields in the forward and backward Schwinger-Keldysh contours, respectively, and use their thermal correlation functions in the Schwinger-Keldysh contour~\cite{Li:2019qkf}.

\subsection{Dynamics of the nucleon spin in a pion gas}
\label{sec:nucleon}

\begin{figure}
 \centering
    \includegraphics[width=0.6\linewidth]{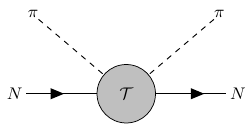}
\cout{
  \scalebox{0.9}{
    \begin{tikzpicture}[baseline=(o.base)]
     \begin{feynhand}
      \vertex (o) at (0,0);
      \vertex (f1) at (2,0) {};
      \vertex (f2) at (-2,0) {};
      \vertex (d1) at (1.65,1.4);
      \vertex (d2) at (-1.65,1.4);
      \node at (-1.75,1.5) {$\pi$};
      \node at (1.75,1.5) {$\pi$};
      \node at (-2.1,0) {$$};
      \node at (-2.1,0) {$N$};
      \node at (2.1,0) {$N$};
      \propag [with arrow = 0.7] (o) to (f1);
      \propag [with arrow = 0.3] (f2) to (o);
      \propag [sca] (d1) to (o);
      \propag [sca] (d2) to (o);
      \filldraw[fill=lightgray] (0,0) circle [radius=5.5mm];
      \vertex at (0,0) {{\small $\matrixT$}};
     \end{feynhand}
   \end{tikzpicture}}
}
 \caption{
 The nucleon-pion scattering that dominantly controls in-medium properties of nucleons up to $T=150\,$MeV.
% in chiral perturbation theory.
 }
 \label{fig-N-pi-scattering}
\end{figure}
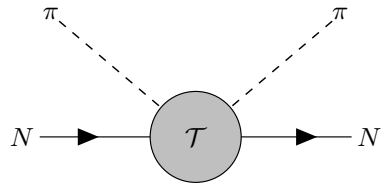

The leading order interaction that we consider in this work is the two-body baryon-pion scattering shown in Fig.~\ref{fig-N-pi-scattering}.
For nucleons, it has been known that this is the dominant process of nucleon-pion interactions for temperatures up to $150$ MeV \cite{Leutwyler:1990uq}.
In the chiral perturbation theory, there are tree-level diagrams composed of $NN\pi$ and $NN\pi\pi$ vertices, but the chiral perturbation theory is no longer justified for temperatures above $100$ MeV.
In addition, the contribution from the $s$-channel $\Delta$-baryon resonance is significant for this temperature range.
Many state-of-the-art models of nucleon-pion scattering \cite{Jenkins:1990jv,Hemmert:1997ye,Fettes:2000bb} include higher baryon resonances, as well as the loop corrections.
The purpose of these models is to use the existing experimental measurements of nucleon-pion scattering cross sections, which in turn can be used to obtain the scattering amplitudes, to constrain the model parameters.

For these reasons, let us first express the evolution equation of the spin density matrix in terms of the nucleon-pion scattering amplitudes in general, which is model-independent.
On the one hand, we may then use the scattering amplitudes that are determined from the experiments, without need to resort to a model calculation.
The existing nucleon-pion scattering data can provide the necessary nucleon-pion scattering amplitudes for the pion gas of temperature up to about $200$ MeV \cite{Leutwyler:1990uq}.
Moreover, any model computation of scattering amplitudes, including chiral perturbation theory, can be used in our formulation of the time evolution equation of the nucleon spin density matrix.

In the following, we assume that the nucleon density matrix is diagonal in the isospin space, and introduce the density matrix for a single nucleon in the isospin space as
\be
\hat{\rho}_N=\left(\begin{array}{cc}\hat{\rho_p} & 0\\0& \hat{\rho_n}\end{array}\right),
\ee
where $\hat\rho_p$ and $\hat\rho_n$ are the density matrices for the isospin $I_z=\pm 1/2$ states, i.e., proton and neutron, respectively.
To describe the effects of the nucleon-pion scattering in the time evolution of the density matrix, we need to expand the unitary time evolution operators, $\hat U(\Delta t)$ and $\hat U^\dagger (\Delta t)$, to at least fourth order of the nucleon-pion interaction Hamiltonian, and compute the thermal correlation functions of the pion operators in Gaussian factorization approximation.
In the limit $\Delta t\gg \tau$, where $\tau$ is the correlation time of the scattering process, we obtain a first-order time evolution equation for the nucleon density matrix $\hat{\rho}_N(t)$, which is expressed in terms of the nucleon-pion scattering amplitude and its complex conjugate.
It is also possible to take a different route to obtain the same result from the collision term of the Kadanoff-Baym equation as shown in Appendix \ref{sec:Kadanoff-Baym}.

\begin{figure*}[htb]
 \centering
\includegraphics[width=0.85\linewidth]{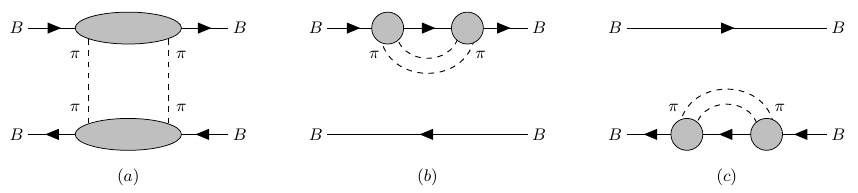}  
\cout{
\scalebox{0.9}{
    \begin{tikzpicture}[baseline=(o.base)]
     \begin{feynhand}
      \vertex (o) at (0,0);
      \vertex (f1) at (2,0) {};
      \vertex (f2) at (-2,0) {};
      \vertex (f3) at (2,2) {};
      \vertex (f4) at (-2,2) {};
      \vertex (o2) at (0,2);
      \vertex (p1) at (0.75,0);
      \vertex (p2) at (-0.75,0);
      \vertex (p3) at (0.75,2);
      \vertex (p4) at (-0.75,2);
      \node at (-1,1.5) {$\pi$};
      \node at (1,1.5) {$\pi$};
      \node at (-1,0.5) {$\pi$};
      \node at (1,0.5) {$\pi$};
      \node at (-2.1,0) {$B$};
      \node at (2.1,0) {$B$};
      \node at (-2.1,2) {$B$};
      \node at (2.1,2) {$B$};
      \node at (0,-0.8) {$(a)$};
      \propag [with arrow = 0.25] (f1) to (o);
      \propag [with arrow = 0.75] (o) to (f2);
      \propag [with arrow = 0.75] (o2) to (f3);
      \propag [with arrow = 0.25] (f4) to (o2);
      \propag [sca] (p1) to (p3);
      \propag [sca] (p2) to (p4);
      \filldraw[fill=lightgray] (0,0) circle (10mm and 3mm);
      \filldraw[fill=lightgray] (0,2) circle (10mm and 3mm);
     \end{feynhand}
   \end{tikzpicture}} 
   \qquad
  \scalebox{0.9}{
    \begin{tikzpicture}[baseline=(o.base)]
     \begin{feynhand}
      \vertex (o) at (0,0);
      \vertex (f1) at (2,0) {};
      \vertex (f2) at (-2,0) {};
      \vertex (f3) at (2,2) {};
      \vertex (f4) at (-2,2) {};
      \vertex (o2) at (0,2);
      \vertex (p1) at (0.9,2);
      \vertex (p2) at (0.6,2);
      \vertex (p3) at (-0.9,2);
      \vertex (p4) at (-0.6,2);
      \node at (-1,1.5) {$\pi$};
      \node at (1,1.5) {$\pi$};
      \node at (-2.1,0) {$B$};
      \node at (2.1,0) {$B$};
      \node at (-2.1,2) {$B$};
      \node at (2.1,2) {$B$};
      \node at (0,-0.8) {$(b)$};
      \propag [with arrow = 0.5] (f4) to (f3);
      \propag [with arrow = 0.5] (f1) to (f2);
      \propag [with arrow = 0.75] (o2) to (f3);
      \propag [with arrow = 0.25] (f4) to (o2);
      \propag [sca] (p1) to [out=270,in=270, looseness=1.618]  (p3);
      \propag [sca] (p2) to [out=270,in=270, looseness=1.618] (p4);
      \filldraw[fill=lightgray] (-0.75,2) circle (3mm and 3mm);
      \filldraw[fill=lightgray] (0.75,2) circle (3mm and 3mm);
     \end{feynhand}
   \end{tikzpicture}} 
   \qquad 
     \scalebox{0.9}{
    \begin{tikzpicture}[baseline=(o.base)]
     \begin{feynhand}
      \vertex (o) at (0,0);
      \vertex (f1) at (2,0) {};
      \vertex (f2) at (-2,0) {};
      \vertex (f3) at (2,2) {};
      \vertex (f4) at (-2,2) {};
      \vertex (o2) at (0,2);
      \vertex (p1) at (0.9,0);
      \vertex (p2) at (0.6,0);
      \vertex (p3) at (-0.9,0);
      \vertex (p4) at (-0.6,0);
      \node at (-1,0.5) {$\pi$};
      \node at (1,0.5) {$\pi$};
      \node at (-2.1,0) {$B$};
      \node at (2.1,0) {$B$};
      \node at (-2.1,2) {$B$};
      \node at (2.1,2) {$B$};
      \node at (0,-0.8) {$(c)$};
      \propag [with arrow = 0.25] (f1) to (o);
      \propag [with arrow = 0.75] (o) to (f2);
      \propag [with arrow = 0.5] (f1) to (f2);
      \propag [with arrow = 0.5] (f4) to (f3);
      \propag [sca] (p1) to [out=90,in=90, looseness=1.618]  (p3);
      \propag [sca] (p2) to [out=90,in=90, looseness=1.618] (p4);
      \filldraw[fill=lightgray] (-0.75,0) circle (3mm and 3mm);
      \filldraw[fill=lightgray] (0.75,0) circle (3mm and 3mm);
     \end{feynhand}
   \end{tikzpicture}} 
   }
 \caption{The cross diagram (a), and the self-energy diagrams, (b) and (c), for the nucleon-pion scattering in the time evolution of nucleon spin density matrix. The blob represents the nucleon-pion scattering amplitude or its complex conjugate.
 }
 \label{fig-Schwinger-Keldysh}
\end{figure*}
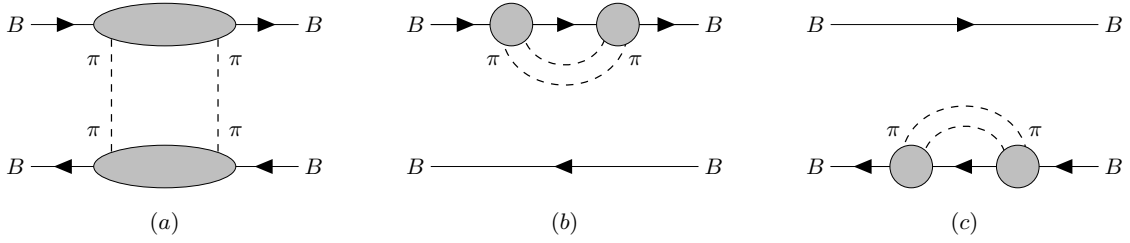

Suppose that the scattering amplitude of a process $N^s_\alpha(\bm p)+\pi^a(\bm k)\to N^{s'}_\beta(\bm p')+\pi^b(\bm k')$ is represented by a matrix, $[\matrixT^{ba}_{\beta\alpha}(\bm p',\bm k';\bm p,\bm k)]_{s',s} = \matrixT^{ba}_{\beta\alpha}(\bm p',s',\bm k';\bm p,s,\bm k)$, in spin space with $s,s'={\pm 1/2}$.
Here, $\alpha,\beta=p,n$ and $a,b=1,2,3$ specify the nucleon and pion isospin states, respectively.
We assume that the pion gas is isospin symmetric.
We then find that the time evolution of the spin density matrix for the proton is given by 
\begin{widetext}
\begin{align}
 \frac{\partial \rho_p^{2\times2} (\bm p)}{\partial t}
 =&\int_{\bm p',\bm k,\bm k'}
 \sum_{a,b,\beta} \matrixT_{p\beta}^{ba}(\bm p,\bm k;\bm p',\bm k')
 \rho_\beta^{2\times2} (\bm p')
 \big[
  \matrixT^{ba}_{p\beta}(\bm p,\bm k;\bm p',\bm k')
 \big]^\dag 
 \nonumber \\
 &\qquad \times 
 n_B(\epsilon_{\bm k'})
 \big[1+n_B(\epsilon_{\bm k})\big]
 (2\pi)^4 \delta^{(3)} 
 (\bm p +\bm k-\bm p'-\bm k')\delta(E_{\bm p}+\epsilon_{\bm k}-E_{\bm p'}-\epsilon_{\bm k'})
%\nonumber\\
%&+\int_{\bm p',\bm k,\bm k'} \sum_{a,b}
%T_{pn}^{ba}(\bm p,\bm k;\bm p',\bm k') 
%\rho_n^{2\times2} (\bm p')
%\big[
%  T^{ba}_{pn}(\bm p,\bm k;\bm p',\bm k')
%\big]^\dag
%\nonumber\\
%&\qquad \times 
%n_B(\epsilon_{\bm k'})
%\big[1+n_B(\epsilon_{\bm k})\big]
%(2\pi)^4 \delta^{(3)} (\bm p+\bm k-\bm p'-\bm k')
%\delta(E_{\bm p}+\epsilon_{\bm k}-E_{\bm p'}-\epsilon_{\bm k'})
\nonumber\\
&-\tilgamma_N (\bm p) \rho_p^{2\times2} (\bm p),
\label{time} 
\end{align}
where $\epsilon_{\bm k}=\sqrt{\bm k^2+m_\pi^2}$ and $E_{\bm p}={\bm p^2/ (2M_N)}$ are the energy dispersion relations of the pions and the nucleons, respectively.
Note that $\rho_{\alpha}(\bm p)$ and $\matrixT^{ba}_{\beta\alpha}(\bm p,\bm k;\bm p',\bm k')$ are $2 \times 2$ matrices in the spin space.
We also introduced the spin and isospin-independent ``thermal damping rate"
$\tilgamma_N(\bm p)$ given by
\begin{align}
 \tilgamma_N (\bm p)
 =& \int_{\bm p',\bm k,\bm k'} 
 \sum_{a,b,\beta}
 \left|\matrixT^{ba}_{\beta p}(\bm p',\bm k';\bm p,\bm k)\right|^2
% \sum_{a,b,\beta}
% \left|T^{ba}_{\beta p}(\bm p',\bm k';\bm p,\bm k)\right|^2
% \nonumber\\
% &\times 
% n_B(\epsilon_{\bm k'})
% \big[1+n_B(\epsilon_{\bm k})\big]
 n_B(\epsilon_{\bm k})
 \big[1+n_B(\epsilon_{\bm k'})\big]
 (2\pi)^4 \delta^{(3)} 
 (\bm p+\bm k-\bm p'-\bm k') 
 \delta(E_{\bm p}+\epsilon_{\bm k}-E_{\bm p'}-\epsilon_{\bm k'}),
 \label{gamma}
\end{align}
\end{widetext}
which is the total scattering rate of a nucleon of momentum $\bm p$ interacting with thermal pions.
The pictorial representation of each contribution is given in Fig.~\ref{fig-Schwinger-Keldysh}. 
The first two terms arise from the ``cross diagram", and the damping rate comes from the self-energy terms (see, e.g., Ref.~\cite{Li:2019qkf}).
The time evolution equation for $\rho_n$ is the same, with the replacement $p\leftrightarrow n$ in all terms. 
Simply speaking, the cross diagram is a product of scattering amplitude and its complex conjugate, but with different initial and final states depending on the components of the density matrix. 
On the other hand, the total self-energy term is simply the absolute square of the scattering amplitude, which represents the usual scattering rate of the nucleon-pion scattering. 
The self-energy term describes the loss of probability from the initial states due to scattering processes, while the cross diagram restores the total probability to make sure that the time evolution is unitary. 
This structure is ubiquitous in the evolution of open quantum systems, and has a form of the Lindblad equation~\cite{Lindblad:1975ef}.

The normalization of the $\matrixT$-matrix elements in the above expression is conventional in the context of nonrelativistic scattering theory, which is related to the nonrelativistic scattering amplitudes as follows.
The thermal damping rate is equal to the total scattering rate of a nucleon with all thermal pions in the background. 
In the heavy nucleon mass limit, the velocity of a nucleon is negligibly small, the final state pion energy is equal to the initial one, and the cross-section depends only on the pion energy. 
In this limit, we have a simplified formula
\be
\tilgamma_N (\bm p) = 
\int_{\bm k} \sigma_{\rm tot}(\epsilon_{\bm k}) v_{\bm k}
n_B(\epsilon_{\bm k})
\big[1+n_B(\epsilon_{\bm k})\big],
\ee
where $v_{\bm k}={|{\bm k}|/\epsilon_{\bm k}}$ is the pion velocity, and $v_{\bm k}n_B(\epsilon_{\bm k})$ is the flux of incoming pions with momentum $\bm k$. 
The $\sigma_{\rm tot}(\epsilon_{\bm k})$ is the total cross section for pion energy $\epsilon_{\bm k}$ with all possible isospin states, i.e., 
\be
\sigma_{\rm tot}=\sum_a \sigma_{p\pi^a\to p\pi^a}+\sigma_{p\pi^0\to n\pi^+}+\sigma_{p\pi^-\to n\pi^0}.
\ee
Comparing with the expression (\ref{gamma}), we find
\begin{align}
 \sigma_{\rm tot}(\epsilon_{\bm k})
&= {2\pi\over v_{\bm k}^2}\int_{\bm k'}
 \sum_{ab,\beta}
 \left|\matrixT^{ba}_{\beta p}(\bm p',\bm k';\bm p,\bm k)\right|^2
 \delta(|\bm k'|-|\bm k|)
 \nonumber \\ 
 &= { \epsilon_{\bm k}^2\over (2\pi)^2}
 \int \diff \Omega' 
 \sum_{a,b,\beta} 
 \left|\matrixT^{ba}_{\beta p}(\bm p',\bm k';\bm p,\bm k)\right|^2
 \nonumber \\ 
 &= \int \diff \Omega' \sum_{a,b,\beta} {\diff\sigma^{ba}_{\beta p}\over \diff\Omega'}.
\end{align}
Here, ${\diff\sigma^{ba}_{\beta p}/ \diff\Omega'}$ is the differential cross section with fixed spin and isospin states ($\int \diff \Omega$ denotes the angular integration). 
From the relation ${\diff\sigma/ \diff\Omega}=|f(\Omega)|^2$, where $f$ is the nonrelativistic scattering amplitude related to the conventionally reported experimental data, we find the normalization relation
\be
f(\bm k',\bm k)= -{\epsilon_{\bm k}\over (2\pi)}\matrixT(\bm k',\bm k).
\ee
Another conventional normalization is the relativistic amplitude given by 
\be 
\mathcal{M}=-(2M_N)(2\epsilon_{\bm k})\matrixT=(8\pi) M_N f.
\label{eq:rela-nonrela}
\ee

As mentioned above, the nucleon velocity is negligible in the heavy nucleon mass limit, $v\sim \sqrt{T/ M_N}\ll 1$, where we used $E_{\bp}={\bp^2/(2M_N)} \sim T$ with temperature~$T$.
The change of nucleon energy is thus small compared to the pion energy, i.e., $\Delta E_N\sim vk \ll k$. 
In this limit, the scattering amplitudes and the evolution equation (\ref{time}) become independent of the nucleon momentum, which allows us to express the evolution equation in terms of the total densities \begin{equation}
    n_\alpha=\int_{\bm p} n_\alpha(\bm p) \quad\mbox{and}\quad \bm S_\alpha=\int_{\bm p} \bm S_\alpha(\bm p)
\end{equation}
 only, where $\alpha=p,n$.
Furthermore, the scattering amplitude is constrained by symmetry. 
The parity, rotational, and isospin symmetry dictates the form of the scattering amplitude as \cite{Fettes:2000bb,Matsinos:1997pb}
\begin{align}
 \matrixT^{ba}_{\beta \alpha}(\bm p',\bm k';\bm p,\bm k)
 &= \delta^{ba} \delta_{\beta\alpha}
 \big[
  {\bf 1}_{2\times 2} g^+
  + \rmi \bm\sigma\cdot(\bm k'\times\bm k) h^+
 \big]
 \nonumber \\
 & +\rmi \epsilon^{bac}(\tau^c)_{\beta\alpha}
 \big[ {\bf 1}_{2\times 2} g^-
 + \rmi \bm\sigma\cdot(\bm k'\times\bm k) h^-
 \big],
\end{align}
where $g^\pm$ and $h^\pm$ are functions of $\epsilon_{\bm k}=\epsilon_{\bm k'}$ and the angle $\theta'$ between $\bm k$ and $\bm k'$, and $({\bf 1}_{2\times 2},\bm\sigma)$ are the matrices in spin space with indices $s',s$, and $\tau^c$ is the Pauli matrix in isospin space.
The first term is the charge-conserving amplitude, and the second term is the charge-exchanging amplitude. 
Similarly, $g^\pm$ and $h^\pm$ are the spin conserving and spin flipping amplitudes, respectively.  

In terms of $(g^\pm,h^\pm)$, the time evolution equation (\ref{time}) can be expressed as follows:
\begin{widetext}
\begin{align}
 \frac{\partial \rho_p^{2\times 2}}{\partial t}
  =& -\int_{\bm k,\bm k'} 
  \left[
   4|g^-|^2+3(|h^+|^2+2|h^-|^2)|\bm k\times\bm k'|^2
  \right]
  n_B(\epsilon_{\bm k}) 
  \big[1+n_B(\epsilon_{\bm k}) \big] 
  2\pi \delta(\epsilon_{\bm k}-\epsilon_{\bm k'})\rho_p^{2\times 2}
  \nonumber\\
  %&+&\int_{\bm k}\int_{\bm k'} (3(|h^+|^2+2|h^-|^2)\left(\bm\sigma\cdot(\bm k\times\bm k') \rho_p^{2\times 2}\bm\sigma\cdot(\bm k\times\bm k')\right)n_B(\epsilon_{\bm k})(1+n_B(\epsilon_{\bm k})) (2\pi)\delta(\epsilon_{\bm k}-\epsilon_{\bm k'})\nonumber\\
  &+ \int_{\bm k,\bm k'} 
  (3|h^+|^2+2|h^-|^2)
  \left[\bm\sigma\cdot(\bm k\times\bm k') \rho_p^{2\times 2}\bm\sigma\cdot(\bm k\times\bm k')\right] n_B(\epsilon_{\bm k})
  \big[1+n_B(\epsilon_{\bm k})\big] 
  2\pi \delta(\epsilon_{\bm k}-\epsilon_{\bm k'})
  \nonumber\\
  &+ 4\int_{\bm k,\bm k'} |g^-|^2n_B(\epsilon_{\bm k})
  \big[1+n_B(\epsilon_{\bm k})\big] 
  2\pi \delta(\epsilon_{\bm k}-\epsilon_{\bm k'})\rho_n^{2\times 2}
  \nonumber\\
  &+ 4\int_{\bm k,\bm k'} |h^-|^2
  \left[
    \bm\sigma\cdot(\bm k\times\bm k') \rho_n^{2\times 2}\bm\sigma\cdot(\bm k\times\bm k')
  \right]
  n_B(\epsilon_{\bm k})
  \big[1+n_B(\epsilon_{\bm k})\big] 
  2\pi \delta(\epsilon_{\bm k}-\epsilon_{\bm k'}).\label{eq:drhodt-gh}
\end{align}
\end{widetext}

Defining 
\begin{math}
\rho_\alpha^{2\times 2}=(1/2)n_\alpha {\bf 1}_{2\times 2}+\bm S_{\alpha}\cdot\bm\sigma 
\end{math}
and taking a trace of equation (\ref{eq:drhodt-gh}) 
multiplied by ${\bf 1}_{2\times 2}$ or $\bm\sigma$, we find, after some algebra, 
that the evolution equation 
for the proton number and spin densities becomes
\be
 {\partial n_p\over\partial t}
 = -\gamma_N (n_p-n_n)\,,\quad {\partial \bm S_p\over\partial t}=-\gamma_1 \bm S_p-\gamma_2 \bm S_n,
\ee
where we introduced the proton-neutron flipping rate $\gamma_N$ given by
\begin{multline}
 \gamma_N
 = 4\int_{\bm k,\bm k'}
 \left(
  |g^-|^2+|h^-|^2|\bm k\times\bm k'|^2
 \right) 
 \\
 \qquad \times
 n_B(\epsilon_{\bm k})
 \big[1+n_B(\epsilon_{\bm k})\big] 
 2\pi \delta (\epsilon_{\bm k}-\epsilon_{\bm k'}),
\end{multline}
and two relaxation rates $\gamma_1$ and $\gamma_2$ given by
\begin{align}
 \gamma_1
 &= 4\int_{\bm k,\bm k'} 
 \left[
  |g^-|^2+\left(|h^+|^2+{5\over 3}|h^-|^2\right)|\bm k\times\bm k'|^2
 \right] 
 \nonumber \\
 &\qquad \times 
 n_B(\epsilon_{\bm k})
 \big[1+n_B(\epsilon_{\bm k})\big] 2\pi
 \delta (\epsilon_{\bm k}-\epsilon_{\bm k'}),
 \end{align}
 \begin{align}
 \gamma_2 
 &=4\int_{\bm k,\bm k'} 
 \left[
  -|g^-|^2+{1\over 3}|h^-|^2|\bm k\times\bm k'|^2
 \right] 
 \nonumber \\
 &\qquad \times 
 n_B(\epsilon_{\bm k}) 
 \big[1+n_B(\epsilon_{\bm k}) \big] 
 2\pi\delta (\epsilon_{\bm k}-\epsilon_{\bm k'}).
\end{align}
The equation for the neutron number and spin densities are the same with the replacement $p\leftrightarrow n$. 
As a result, we see that 
$\gamma_{N_s} := \gamma_1+\gamma_2$ corresponds to the spin relaxation rate of the total nucleon spin density, $\bm S_N =\bm S_p+\bm S_n$, i.e., 
\be
 {\partial \bm S_N\over\partial t}=-\gamma_{N_s} \bm S_N.
\ee
We then find a formula for the nucleon spin relaxation rate $\gamma_{N_s}$ in terms of 
the scattering amplitude as
\begin{align}
 \gamma_{N_s} 
 &= 4\int_{\bm k,\bm k'} 
 \left( |h^+|^2+2|h^-|^2 \right) |\bm k\times\bm k'|^2
 \nonumber \\
 &\qquad \times
 n_B(\epsilon_{\bm k}) 
 \big[1+n_B(\epsilon_{\bm k}) \big]
 2 \pi \delta (\epsilon_{\bm k}-\epsilon_{\bm k'}).
\label{eq:spin-relaxation-nucleon}
\end{align}
Note that the spin relaxation rate is determined by the spin-flipping amplitudes, 
i.e., $h^\pm$, as expected.
Although it is not of our main interest, the nucleon thermal damping rate $\tilgamma_N$ defined in Eq.~\eqref{gamma}
is given by
\begin{align}
 \tilgamma_N 
 &= 3\int_{\bm k,\bm k'}
 \hspace{-4pt}
 \left[|g^+|^2+2|g^-|^2+ 
 \left(|h^+|^2+2|h^-|^2\right)
 |\bm k\times\bm k'|^2
 \right]
 \nonumber \\
 &\qquad \times
 n_B (\epsilon_{\bm k}) 
 \big[1+n_B(\epsilon_{\bm k}) \big] 
 2\pi \delta (\epsilon_{\bm k}-\epsilon_{\bm k'}).
\end{align}

\subsection{Dynamics of the \texorpdfstring{$\Lambda$}{Lambda}-baryon spin in a pion gas}
\label{sec:Lambda}

One advantage of our formulation is that we can study the time evolution of the spin density matrix of $\Lambda$ baryons in the same framework. 
The $\Lambda$ baryon is an isospin singlet state, and has spin $1/2$ and parity $P=+1$. 
The scattering process with the $\Lambda$ baryon and the pion, $\Lambda(s)+\pi^a(\bm k)\to \Lambda(s')+\pi^b(\bm k')$, has only the isospin conserving amplitude.
Thus, we can parametrize the corresponding $\matrixT$-matrix as
\be
 \matrixT^{ba} (\bm k';\bm k)
 = \delta^{ab}
  \big[
   {\bf 1}_{2\times 2} g
   +\rmi \bm\sigma\cdot(\bm k'\times\bm k) h
  \big].
  \label{eq:T-matrix-lambda-pi}
\ee
We then derive the equation of motion for the spin density matrix, $\rho_\Lambda^{2\times2} = \left( {\bf 1}_{2\times 2} n_\Lambda/2+\bm\sigma\cdot\bm S_\Lambda \right)_{ss'}$
with the number density $n_\Lambda$ and spin densiy $\bm{S}_\Lambda$ for the $\Lambda$ baryon, as
\begin{align}
 \frac{\partial \rho_\Lambda^{2\times 2}}{\partial t}
 =& \int_{\bm k,\bm k'} 
 \sum_{a,b} \matrixT^{ba} (\bm k;\bm k') 
 \rho_\Lambda^{2\times 2}
 \big[\matrixT^{ba}(\bm k;\bm k') \big]^\dag
 \nonumber \\
 &\quad \times 
 n_B(\epsilon_{\bm k'})
 \big[1+n_B(\epsilon_{\bm k}) \big]
 2\pi \delta(\epsilon_{\bm k}-\epsilon_{\bm k'})
 \nonumber \\
 &- \tilgamma_{\Lambda} \rho_\Lambda^{2\times 2},
 \label{lambdatime}
\end{align}
where the thermal damping rate $\tilgamma_\Lambda$ is given by
\begin{align}
 \tilgamma_{\Lambda} 
 &= \int_{\bm k,\bm k'}
 \sum_{a,b} \left|\matrixT^{ba}(\bm k';\bm k)\right|^2
 \nonumber \\
 &\qquad \times 
 n_B(\epsilon_{\bm k})
 \big[1+n_B(\epsilon_{\bm k'})\big]
 2\pi \delta(\epsilon_{\bm k}-\epsilon_{\bm k'}).
 \label{lambdagamma}
\end{align}
Multiplying  Eq.~\eqref{lambdatime} by ${\bf 1}_{2\times 2}$ or $\bm\sigma$ 
and taking a trace, we obtain the evolution equation for 
$n_\Lambda$ and $\bm{S}_\Lambda$ in the form
\be
 {\partial n_{\Lambda} \over\partial t} =0,
 \quad 
 {\partial \bm{S}_\Lambda \over\partial t}
 = - \gamma_{\Lambda_s} \bm{S}_\Lambda.
\ee
Here, the spin relaxation rate, $ \gamma_{\Lambda_s}$, is expressed in terms of $(g,h)$ as
\begin{align}
 \gamma_{\Lambda_s}
 &= 4 \int_{\bm k,\bm k'} |h|^2
 |\bm k\times\bm k'|^2 
 \nonumber \\
 &\qquad \times
 n_B(\epsilon_{\bm k})
 \big[1+n_B(\epsilon_{\bm k})\big] 
 2\pi\delta (\epsilon_{\bm k}-\epsilon_{\bm k'}).
% \nonumber \\
% &= \frac{4}{3\pi^3}
% \int_{m_{\pi}}^\infty \diff \omega \,
% \omega^2 (\omega^2 - m_\pi^2)^3 |h|^2
% n_B(\omega) \big[1+n_B(\omega)\big],
 \label{eq:spin-relaxation-Lambda}
\end{align}
Although it is not of our main interest, the thermal damping rate is expressed as
\begin{align}
 \tilgamma_{\Lambda} 
 &= 3 \int_{\bm k,\bm k'} 
 \left(|g|^2+ |h|^2|\bm k\times\bm k'|^2\right)
 \nonumber \\
  &\qquad \times 
  n_B(\epsilon_{\bm k})
  \big[1+n_B(\epsilon_{\bm k})\big] 
  2\pi \delta (\epsilon_{\bm k}-\epsilon_{\bm k'}).
 \label{eq:thermal-relaxation-Lambda}
\end{align}

\section{Evaluation of the spin relaxation rate of baryons}
\label{sec:spin-relaxation}

In this section, we evaluate the baryon spin relaxation rates based on the formulae \eqref{eq:spin-relaxation-nucleon} and \eqref{eq:spin-relaxation-Lambda}, which express 
the spin relaxation rate in terms of the baryon-pion scattering amplitude.
In Sec.~\ref{sec:nucleon-spin-relaxation}, we apply the $s$-channel $\Delta$-resonanance approximation, and 
use the available experimental data on the nucleon-pion scattering amplitudes expressed in terms of the scattering phase shift, to compute the nucleon spin relaxation rate. 
In Sec.~\ref{sec:Lambda-spin-relaxation},
we introduce reasonable phenomenological models for the $\Lambda\pi$ scattering amplitude, utilizing available experimental data on strange baryons, as well as the chiral perturbation theory at sufficiently low energy, and compute the spin relaxation rate of $\Lambda$-baryons.

\subsection{The spin relaxation rate of nucleons in \texorpdfstring{$\Delta$}{Delta}-resonance approximation}
\label{sec:nucleon-spin-relaxation}

\begin{figure*}
\centering
\includegraphics[width=0.6\linewidth]{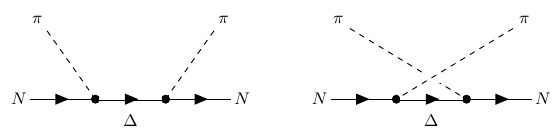}  
\cout{
 \scalebox{0.9}{
    \begin{tikzpicture}[baseline=(o.base)]
     \begin{feynhand}
      \vertex (o) at (0,0) {};
      \vertex (f1) at (2,0) {};
      \vertex (f2) at (-2,0) {};
      \vertex (v1) [dot] at (0.66,0) {};
      \vertex (v2) [dot] at (-0.66,0) {};
      \vertex (d1) at (1.65,1.4) {};
      \vertex (d2) at (-1.65,1.4) {};
      \node at (-1.75,1.5) {$\pi$};
      \node at (1.75,1.5) {$\pi$};
      \node at (-2.1,0) {$N$};
      \node at (2.1,0) {$N$};
      \node at (0,-0.4) {$\Delta$};
      \propag [with arrow = 0.5] (v1) to (f1);
      \propag [double, double distance = 1pt, with arrow = 0.5] (v2) to (v1);
      \propag [with arrow = 0.5] (f2) to (v2);
      \propag [sca] (d1) to (v1);
      \propag [sca] (d2) to (v2);
     \end{feynhand}
    \end{tikzpicture}} 
 \qquad 
  \scalebox{0.9}{
     \begin{tikzpicture}[baseline=(o.base)]
      \begin{feynhand}
       \vertex (o) at (0,0) {};
       \vertex (f1) at (2,0) {};
       \vertex (f2) at (-2,0) {};
       \vertex (v1) [dot] at (0.66,0) {};
       \vertex (v2) [dot] at (-0.66,0) {};
       \vertex (d1) at (1.65,1.4) {};
       \vertex (d2) at (-1.65,1.4) {};
       \node at (-1.75,1.5) {$\pi$};
       \node at (1.75,1.5) {$\pi$};
       \node at (-2.1,0) {$N$};
       \node at (2.1,0) {$N$};
       \node at (0,-0.4) {$\Delta$};
       \propag [sca] (d2) to (v1);
       \propag [sca, top] (d1) to (v2);
       \propag [with arrow = 0.5] (v1) to (f1);
       \propag [double, double distance = 1pt, with arrow = 0.5] (v2) to (v1);
       \propag [with arrow = 0.5] (f2) to (v2);
      \end{feynhand}
   \end{tikzpicture}}
   }
 \caption{The tree-level Feynman diagrams for the amplitude of nucleon-pion scattering with $\Delta$-baryon resonance.}
 \label{fig-Delta}
\end{figure*}
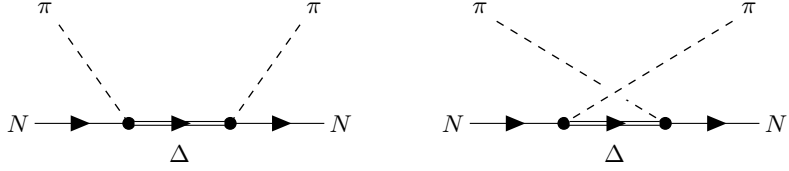

It has been known that $\Delta$-baryon---the $s$-channel resonance of the spin $3/2$ and isospin $3/2$---gives the dominant contribution to the nucleon damping rate in the temperature range up to 150 MeV \cite{Leutwyler:1990uq}. 
Based on this, it is reasonable to compute the spin relaxation rate in  the $\Delta$-resonance approximation. 
The tree-level diagrams for the amplitude with an intermediate $\Delta$-baryon are depicted in Fig.~\ref{fig-Delta}, where only the first diagram produces the $s$-channel resonance behavior.
However, instead of theoretically evaluating the scattering amplitude, we are going to use the experimental data on the resonant scattering amplitude.
This analysis has the advantage that the resulting prediction will be model-independent.

We first express the scattering amplitude assuming that the $\Delta$-baryon resonance gives a dominant contribution.
Since $\Delta$-baryons are spin and isospin $3/2$ states, a resonance is seen only in the particular channel in partial wave analysis of the nucleon-pion scattering, i.e., the channel where the total angular momentum $J$ and the isospin $I$ are both $3/2$.
Here we note that the amplitude for the $I=3/2$ states is given by the combination $(g_{3/2},h_{3/2})=(g^+-g^-,h^+-h^-)$, whereas for the $I=1/2$ state by $(g_{1/2},h_{1/2})=(g^++2g^-,h^++2h^-)$.%
\footnote{This can be seen by observing that the isospin exchange operator $\rmi\epsilon^{bac} (\tau^c)_{\beta\alpha}=-2(I_\pi^cI_N^c)^{ba}_{\beta\alpha}$, where $I_\pi^c$ and $I_N^c$ are the isovector components of the pion and nuleon isospin operators respectively ($I_\pi=1$ and $I_N=1/2$). The eigenvalues of this operator are equal to $11/4-I(I+1)$ for total isospin $I$. } 
Since the dominant resonance contribution appears only in the $I=3/2$ channel, we approximate $(g_{1/2},h_{1/2})\approx 0$, which then gives us $(g_{3/2},h_{3/2})\approx -3(g^-,h^-) \approx (3/ 2)(g^+,h^+)$ for the resonant amplitude.
Then, the spin-damping rate for nucleons is expressed in terms of the $I=3/2$ amplitudes as
\begin{align} 
 \gamma_{N_s}
 &\approx {8\over 3}\int_{\bm k,\bm k'}|h_{3/2}|^2|\bm k\times\bm k'|^2
 \nonumber \\
 &\quad \times 
 n_B(\epsilon_{\bm k}) 
 \big[1+n_B(\epsilon_{\bm k})\big] 
 2\pi\delta (\epsilon_{\bm k}-\epsilon_{\bm k'})
 \label{deltaspin}.
\end{align}
Similarly, the nucleon flipping and the thermal damping rates are given by
\begin{align}
 \gamma_N 
 &\approx 
 {4\over 9} \int_{\bm k,\bm k'}
  \left(
    |g_{3/2}|^2+|h_{3/2}|^2|\bm k\times\bm k'|^2
  \right)
 \nonumber \\
 &\quad \times 
 n_B(\epsilon_{\bm k}) \big[1+n_B(\epsilon_{\bm k})\big] 
 2 \pi \delta (\epsilon_{\bm k}-\epsilon_{\bm k'}) , 
 \nonumber \\
 \tilgamma_N 
 &\approx 
 2 \int_{\bm k,\bm k'}
 \left(
  |g_{3/2}|^2+|h_{3/2}|^2|\bm k\times\bm k'|^2
 \right)
 \nonumber \\
 &\quad \times 
 n_B(\epsilon_{\bm k}) \big[1+n_B(\epsilon_{\bm k})\big] 
 2\pi \delta (\epsilon_{\bm k}-\epsilon_{\bm k'})  .
\end{align}

The amplitude for the nucleon-pion scattering has been measured in experiments \cite{Hohler:1977em}, which is available in the form of scattering phase shift $\delta(\epsilon_{\bm k})$ \cite{Koch:1980ay}. 
Thus, we need to relate the resonance phase shift measured in experiments with the amplitudes $(g_{3/2},h_{3/2})$ based on the scattering theory.

First of all, rotational symmetry allows us to choose the incoming pion momentum to be $\bm k=k\hat{\bm z}$ while the outgoing one $\bm k'$ is expressed in spherical coordinates system as $\bm k'=k(\sin\theta\cos\phi,\sin\theta\sin\phi,\cos\theta)$.
Then, consider the incoming plane wave state with a particular nucleon spin orientation, say $|s_z=1/2\rangle$, i.e., 
\begin{align}
 \psi_\mathrm{in}
 &= \rme^{\rmi kz} \chi_s(1/2)
 \nonumber \\
 &= \sum_{l=0}^\infty \rmi^l (2l+1)
 j_l(kr) P_l(\cos\theta) \chi_s(1/2),
\end{align}
with the spherical Bessel function $j_l(kr)$, the Legendre polynomial $P_l(\cos\theta)$, and the spin wave function $\chi_s(s')=\langle s|s_z=s'\rangle$.
The resonance appears in the total angular momentum $J=l\pm1/2=3/2$ channel, with two possible values of orbital angular momentum $l=1$ or $l=2$. 
We focus on the $l=1$ state because only the $l=1$ (P) channel exhibits resonance experimentally.
Noting $P_1 (\cos \theta) = \sqrt{{4\pi}/{3}} Y_{1}^{0} (\theta,\phi)$ with $Y_l^m(\theta,\phi)=\langle\theta,\phi|l,m\rangle$, the spherical harmonics of the orbital angular momentum $(l,m)$, we can express the $l=1$ state as
\begin{equation}
 \psi_\mathrm{in}
%  = 3 \rmi j_1(kr) \cos\theta 
%  |s_z=1/2\rangle
%  = 3\rmi j_1(kr) \sqrt{4\pi\over 3}Y_1^0(\theta,\phi)|s_z=1/2\rangle
 = 3\rmi j_1(kr)\sqrt{4\pi\over 3}\langle\theta,\phi|l=1,m=0\rangle\langle s|s_z=1/2\rangle.
\end{equation}

The state $|l=1,m=0\rangle|s_z=1/2\rangle$ in the incoming wave is a linear combination of $J=3/2$ and $J=1/2$ states, i.e., using the Clebsch-Gordan coefficients, we have
\begin{align}
 &|l=1,m=0\rangle|s_z=1/2\rangle
 \nonumber \\
 =& \sqrt{2\over 3}|J=3/2,J_z=1/2\rangle-\sqrt{1\over 3}|J=1/2,J_z=1/2\rangle.
\end{align}
Note that only the $J=3/2$ wave function is affected by the resonance phase shift, and we neglect the small phase shift in the $J=1/2$ channel. 
From the asymptotic form of the $l=1$ spherical Bessel function 
$j_1(kr) \to -(\rme^{\rmi kr} + \rme^{-\rmi kr}){/(2kr)}$ at $kr \to \infty$, we have the scattering wave that contains only $J=3/2$ state
\begin{align}
 \psi_\mathrm{scatt} 
 &= f(\Omega){\rme^{\rmi kr}\over r}
 \nonumber \\
 &\equiv 
 {3\over 2\rmi k}\sqrt{4\pi\over 3}\sqrt{2\over 3}(\rme^{2\rmi\delta(k)}-1)
 \nonumber \\
 &\quad \times
 \langle\theta,\phi|\langle s |J=3/2,J_z=1/2\rangle{e^{ikr}\over r},
\end{align}
where $\delta(k)=\delta(\epsilon_k)$ is the phase shift in the $J=3/2$ channel.
We then use the inverse of Clebsch-Gordan coefficients to express the scattering wave in the original orbital-spin basis as
\begin{align}
 |J=3/2,J_z=1/2\rangle
 =& \sqrt{2\over 3}|l=1,m=0,s_z=1/2\rangle
 \nonumber \\ 
 +&
 \sqrt{1\over 3}|l=1,m=1,s_z=-1/2\rangle.
  % \nonumber \\
  % &= \sqrt{2\over 3}Y_1^0(\theta,\phi)|s_z=1/2\rangle+\sqrt{1\over 3}Y_1^1(\theta,\phi)|s_z=-1/2\rangle,
\end{align}
Recalling that the initial nucleon spin state is $|s_z=1/2\rangle$, and the final spin state is what appears in the scattering wave, we read off the spin-dependent scattering amplitudes $f(s',s)$ as 
\begin{align}
 f(1/2,1/2)
 &= {3\over 2\rmi k}\sqrt{4\pi\over 3}{2\over 3}(\rme^{2\rmi \delta(k)}-1) Y_1^0(\theta,\phi)
 \nonumber \\
 &= {1\over \rmi k}(\rme^{2\rmi\delta(k)}-1)\cos\theta
 \nonumber \\
 f(-1/2,1/2)
 &={3\over 2\rmi k}\sqrt{4\pi\over 3}{\sqrt{2}\over 3}(\rme^{2\rmi\delta(k)}-1)Y_1^1(\theta,\phi)
 \nonumber \\
 &= -{1\over 2\rmi k}(\rme^{2\rmi\delta(k)}-1)\sin\theta e^{\rmi\phi},
 \label{eq:f-from-plus-1/2}
\end{align}
where we used the spherical harmonics $Y_l^m(\theta,\phi)=\langle\theta,\phi|l,m\rangle$ with $Y_1^{\pm 1}=\mp\sqrt{3/(8\pi)}\sin\theta \exp({\pm \rmi \phi})$ and $Y_1^0=\sqrt{3/ (4\pi)}\cos\theta$.
Repeating the same analysis with the initial spin state $|s_z=-1/2\rangle$, we arrive at
\begin{align}
 f(1/2,-1/2)
 &= {1\over 2\rmi k}(\rme^{2\rmi\delta(k)}-1)\sin\theta \rme^{-\rmi\phi},
 \nonumber \\
 f(-1/2,-1/2) &= f(1/2,1/2).
 \label{eq:f-from-minus-1/2}
\end{align}
We then compare the result for the scattering amplitude in Eqs.~\eqref{eq:f-from-plus-1/2}-\eqref{eq:f-from-minus-1/2} with the expression
\begin{align}
 f(s',s) 
 &=-{\epsilon_{\bm k}\over 2\pi}T(s',s)
 \nonumber \\ 
 &=-{\epsilon_{\bm k}\over 2\pi}
 \big[
  {\bf 1}_{2\times 2} g_{3/2} 
  +\rmi\bm\sigma\cdot(\bm k'\times \bm k)h_{3/2}
  \big]_{s',s}
 \nonumber \\ 
 &= -{\epsilon_{\bm k}\over 2\pi}
 \begin{pmatrix}
  g_{3/2} & -k^2\sin\theta \rme^{-\rmi\phi}h_{3/2} \\ 
  k^2 \sin\theta \rme^{\rmi\phi}h_{3/2} & g_{3/2}
 \end{pmatrix}_{s',s}.
\end{align}
From this, we obtain the relation between the phase shift and the amplitudes for the resonance as follows:
\begin{align}
 g_{3/2} &= -{2\pi\over\epsilon_{\bm k}}{1\over \rmi k}(\rme^{2\rmi\delta(k)}-1)\cos\theta,
 \nonumber \\
 h_{3/2} &= {2\pi\over\epsilon_{\bm k}}{1\over 2\rmi k^3}(\rme^{2\rmi\delta(k)}-1).
\end{align}
Using this in Eq.~\eqref{deltaspin} and performing angular integrations, we obtain the nucleon spin relaxation rate in terms of the phase shifts as
\begin{align}
 \gamma_{N_s} 
 &= {8\over 9\pi}\int_0^\infty \diff k {k\over\epsilon_k}|\rme^{2\rmi\delta(k)}-1|^2 
 n_B(\epsilon_k) \big[1+n_B(\epsilon_k) \big]
 \nonumber \\
 &= {8\over 9\pi} \int_{m_{\pi}}^\infty \diff \omega 
 |\rme^{2\rmi\delta(\omega)}-1|^2 
 n_B(\omega) \big[1+n_B(\omega)\big],
 \label{finres}
\end{align}
where we performed the change of variable from $k=|\bm k|$ to the pion energy $\omega\equiv\epsilon_{k}$. Similarly, we obtain the nucleon flipping and the thermal damping rates as
 $\gamma_N=({1/ 2})\gamma_{N_s}$ and $\tilgamma_N = ({9/ 4})\gamma_{N_s}$ in the $\Delta$-resonance approximation.
 
\begin{figure}
\includegraphics[width=0.95\linewidth]{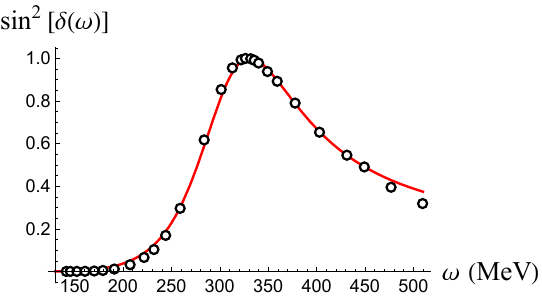}
 \caption{The phase shift of the resonant $I=J=3/2$ channel in the nucleon-pion scattering. 
 The open circles are the experimental data from Ref.~\cite{Koch:1980ay}, and the solid line is our fitting curve given in Eq.~(\ref{fit}).}\label{fig4}
\end{figure}

Relying on Eq.~\eqref{finres}, let us evaluate the spin relaxation rate of nucleons, using the expermentally measured phase shift $\delta(\omega)$.
The scattering phase shift data up to $\omega=500$ MeV in the $I=J=3/2$ channel can be found in Ref.~\cite{Koch:1980ay} (the $P_{33}$ channel in their notation). 
We use these data to model the phase shift as a function of $\omega$ in the simplest Breit-Wigner resonance form:
\be
 \tan \big(\delta(\omega)-\delta_B \big)
 = {\Gamma/2\over \omega_0-\omega}.
 \label{fit}
\ee
We find a good fit with $\Gamma=129$ MeV, $\omega_0=307$ MeV, and the background phase shift $\delta_B=-\pi/9$ (see Fig.~\ref{fig4} for the comparison with the fitting curve and data).
We note that the shape of the phase shift for the energies greater than $500$ MeV is not important in our result for $\gamma_{N_s}$ in the temperature range $T\lesssim 200$ MeV, due to the thermal suppression from the Bose distribution $n_B(\omega)$ in Eq.~(\ref{finres}).

\begin{figure}
\includegraphics[width=0.95\linewidth]{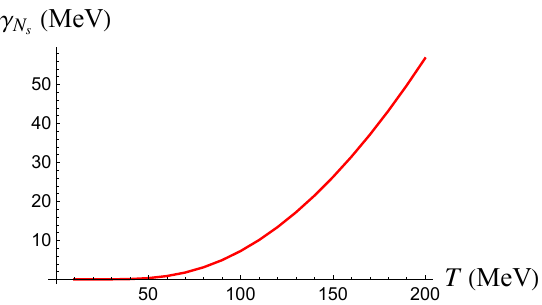}
\caption{The nucleon spin relaxation rate $\gamma_{N_s}$ in the $\Delta$-baryon resonance approximation. }
\label{fig5}
\end{figure}

Substituting this phase shift into Eq.~\eqref{finres}, we numerically evaluate the nucleon spin relaxation rate $\gamma_{N_s}$ as a function of temperature $T$. 
The result is shown in Fig.~\ref{fig5}. 
We see that $\gamma_{N_s}$ is a monotonically increasing function of $T$, and it reaches to values about 
$50$ MeV ($\simeq 1/4$ fm$^{-1}$) for $T$ close to $200$ MeV.

\subsection{The spin relaxation rate of \texorpdfstring{$\Lambda$}{Lambda}-baryons }
\label{sec:Lambda-spin-relaxation}

In Sec.~\ref{sec:Lambda}, we have derived the formula \eqref{eq:spin-relaxation-Lambda} expressing the spin relaxation rate of the $\Lambda$ baryons in terms of the scattering amplitudes $(g,h)$.
Since there is no direct experimental data on $(g,h)$, or equivalent scattering phase shifts, we need to rely on a reasonable theoretical model.
In the following, we consider two approaches: one based on the chiral perturbation theory, and the other relying on the low-energy parametrization of the scattering amplitude with a reasonable value for the scattering length.

\subsubsection{Chiral perturbation theory}

The first approach is the chiral perturbation theory for baryons with $N_f = 3$~\cite{Weinberg:1978kz,Gasser:1983yg,Gasser:1984gg,Scherer-Schindler2011}.
Within the chiral perturbation theory, one can compute the scattering amplitude \eqref{eq:T-matrix-lambda-pi} in the heavy-baryon limit as follows (see Appendix \ref{sec:chiPT} for the derivation):
\begin{align}
 g &\simeq 
 - \frac{2 \bar{m} D^2}{3f_\pi^2 \epsilon_{\bk}}
   \bigg[
   1
   + \frac{\bar{m} \Delta m }{s - m_\Sigma^2} 
   + \frac{\bar{m} \Delta m }{u - m_\Sigma^2}
  \nonumber \\
  &\hspace{54pt}
  - \frac{\bar{m}}{4m_\Lambda} 
   \left( 
   \frac{s-u}{s - m_\Sigma^2}
   - \frac{s-u}{u - m_\Sigma^2}
   \right)
  \bigg],
 \nonumber \\
 h &\simeq 
 \frac{\bar{m}^2 D^2}{3f_\pi^2 m_\Lambda \epsilon_{\bk}}
 \bigg[ 
   \frac{1}{s - m_\Sigma^2}
   - \frac{1}{u - m_\Sigma^2}
 \bigg],
 \label{eq:g-h-chiPT}
\end{align}
where we introduced 
$2\bar{m} \equiv m_\Sigma + m_\Lambda$, $\Delta m \equiv m_\Sigma - m_\Lambda$, and a $D$-type baryon-meson coupling constant $D = 0.8$ which is determined from a semileptonic decay process~\cite{Borasoy:1998pe}.
We also introduced two Mandelstam variables $s$ and $u$, which becomes
$s \simeq (m_\Lambda + E_{\bk} + \epsilon_{\bk})^2$ and 
$u \simeq (m_\Lambda + E_{\bk} - \epsilon_{\bk})^2 - 2 \bk^2 (1+\cos \theta)$
in the center-of-mass frame in the heavy-baryon limit.
Here, we utilized the fact that for elastic scattering in the center-of-mass frame, the absolute values of the incoming and outgoing momenta are the same, i.e., $|\bm{k}|=|\bm{k}'|$.

Due to rotational symmetry, $h$ depends only on $\bk^2$ and $z \equiv \cos \theta$ 
with the angle $\theta$ between $\bk$ and $\bk'$, and we perform a part of the integral in Eq.~\eqref{eq:spin-relaxation-Lambda},  
introduce a change of variable from $k$ to $\omega \equiv \epsilon_{\bk}$
to obtain 
\begin{align}
 \gamma_{\Lambda_s}
 &= \frac{1}{\pi^3}
 \int_{m_\pi}^\infty \diff \omega \omega^2 (\omega^2 - m_\pi^2)^3
 \int_{-1}^1 \diff z 
 \nonumber \\
 &\qquad \times
 |h|^2 
 (1-z^2)
 n_B(\omega)
 \big[1+n_B(\omega)\big] .
\end{align}
Substituting the expression for $h$ in (\ref{eq:g-h-chiPT}) into this equation, and performing the remaining integral, we can evaluate the $\Lambda$-baryon spin relaxation rate numerically.
With a choice of parameters
$m_\pi = 140\,$MeV, $f_\pi = 90\,$MeV, 
$m_\Lambda = 1116\,$MeV, 
$m_\Sigma = 1190\,$MeV, and $D=0.8$, 
we obtain the result (black dashed curve) shown in Fig.~\ref{fig7}.

\subsubsection{Low-energy scattering theory}

Similarly to the $\Delta$-resonance in the $N\pi$ scattering, there is a resonance $\Sigma^*$ appearing in the $\Lambda\pi$ scattering, whose contribution is likely to be large in the temperature window under consideration.
Thus, it is important to include a contribution from $\Sigma^*$-baryon resonance in the evaluation of the $\Lambda$ spin relaxation rate.
In the following, after rewriting the spin relaxation rate in terms of phase shifts, we will employ the low-energy parametrization of the scattering amplitude with a reasonable value of the scattering length.

Let us first recall the relevant facts about hyperons.
First of all, note that there are two strange-baryons with isospin $I=1$, with their masses close to the sum of $\Lambda$ baryon mass, $m_\Lambda=1116$ MeV, and the pion mass, i.e., $m_\Lambda+m_\pi=1256$ MeV.
They are spin 1/2, $\Sigma$ baryon of average mass $m_\Sigma=1190$ MeV, and the spin 3/2, $\Sigma^*(1385)$ baryon, respectively. 
The former is a good candidate for a near threshold bound state of $\Lambda\pi$ system, and the latter is a resonance of a finite width $\Gamma=36$ MeV. 

We note that all of $\Sigma$, $\Sigma^*$, and $\Lambda$ baryons have the parity $P=+1$ while pions have the parity $P=-1$.
Thus, in order to have $\Sigma$ and $\Sigma^*$ baryons as possible intermediate states, the orbital angular momentum $l$ of $\Lambda\pi$ scattering needs to be an odd integer. 
Indeed, considering their spins, we see that both $\Sigma$ and $\Sigma^*$ baryons can only appear in the $l=1$ channel of the $\Lambda\pi$ scattering. 
Specifically, $\Sigma$ baryon can be interpreted as a near threshold bound state in the total angular momentum $J=1/2$ channel, and $\Sigma(1385)$ a resonance in the $J=3/2$ channel, where $J=L+S$ with $l=1$ and $s=1/2$.

In the following analysis, we assume that the scattering phase shift is dominated by these intermediate baryon states, and we approximate them by using the simplest model of phase shifts for a near threshold bound state and a resonance, respectively.  
We note that $\Sigma^*(1385)$ decays dominantly to $\Lambda\pi$ with branching ratio $90\%$, but the remaining $10\%$ goes to $\Sigma\pi$, which means that $\Lambda$ baryon can become $\Sigma$ baryon by scattering with pions. We will neglect this transition to leading approximation, and focus on the $\Lambda$ baryons only. A more complete framework including both $\Lambda$ and $\Sigma$ in the time evolution of their spin density matrix is postponed to a future study. 
Note that both $\Lambda$ and $\Sigma$ have the decay lifetime, $\tau\sim 10^{-10}$s, due to weak interactions, which is much longer than the time scale of QCD plasma created in heavy-ion collisions, and we can treat them as stable particles to a good approximation.%
\footnote{The lifetime of $\Sigma^0$, $\tau\sim 10^{-20}$ s, is shorter than that of $\Sigma^\pm$, by isospin breaking electromagnetic interaction, $\Sigma^0\to \Lambda+\gamma$, but it is still much longer than the time scale of QCD plasma, i.e., $10^{-22}$ s.}

The relation between the scattering phase shifts in $J=1/2$ and $J=3/2$ channels and the amplitudes $(g,h)$ is derived in a similar manner as 
 in the case of nucleons. 
The incoming plane wave of a pion in $l=1$ channel with a $\Lambda$ baryon with spin $s_z=1/2$ is given by the following linear combination of $J=1/2$ and $J=3/2$ states:
\begin{widetext}
\begin{align}
 \psi_{\mathrm{in}}
%  \rme^{\rmi kz} |s_z=1/2\rangle
 &\sim {3\over 2\rmi k}{1\over r}
 (\rme^{\rmi kr} + \rme^{-\rmi kr}) P_1(\cos\theta)
 \chi_s(1/2)
 \nonumber \\
 &= {3\over 2\rmi k}{1\over r} (\rme^{\rmi kr} + \rme^{-\rmi kr})
 \sqrt{4\pi\over 3}
 \langle \theta,\phi|\langle s|\left(\sqrt{2\over 3}|J=3/2,J_z=1/2\rangle-\sqrt{1\over 3}|J=1/2,J_z=1/2\rangle \right).
\end{align}
On the other hand, the out-going scattering wave is given in terms of the two phase shifts, $\delta_{3/2}$ and $\delta_{1/2}$, for $J=3/2$ and $J=1/2$ channels, respectively, as
\begin{align}
 \psi_\mathrm{scatt}
 &= {\sqrt{4\pi}\over 2\rmi k}
\langle \theta,\phi|\langle s| \Big[ \sqrt{2}
  ( \rme^{2\rmi\delta_{3/2}}-1 )
  |J=3/2,J_z=1/2\rangle
  - (\rme^{2\rmi\delta_{1/2}}-1 )
  |J=1/2,J_z=1/2\rangle
 \Big] 
 {\rme^{\rmi kr}\over r}.
\end{align}
Going back to the orbital-spin basis by using
\begin{align}
 |J=3/2,J_z=1/2\rangle
 =& \sqrt{2\over 3}Y_1^0(\theta,\phi)|s_z=1/2\rangle+\sqrt{1\over 3}Y_1^1(\theta,\phi)|s_z=-1/2\rangle,
 \nonumber\\
 |J=1/2,J_z=1/2\rangle
 =& -\sqrt{1\over 3}Y_1^0(\theta,\phi)|s_z=1/2\rangle+\sqrt{2\over 3}Y_1^1(\theta,\phi)|s_z=-1/2\rangle,
\end{align}
\end{widetext}
we find the spin-dependent scattering amplitudes as
\begin{align}
 f(1/2,1/2) 
 &= {1\over 2\rmi k}
 (2\rme^{2\rmi\delta_{3/2}}+\rme^{2\rmi\delta_{1/2}}-3) \cos\theta,
 \nonumber \\
 f(-1/2,1/2)
 &= {1\over 2ik} 
 (-\rme^{2\rmi\delta_{3/2}}+\rme^{2\rmi\delta_{1/2}})\sin\theta \rme^{\rmi\phi}.
\end{align}
Comparing these result with $f(s',s)=-(\epsilon_{\bm k}/ (2\pi)) T(s',s)$, we find the relations
\begin{align}
 g &= -{2\pi\over\epsilon_{\bm k}}{1\over 2\rmi k}
 (2\rme^{2\rmi\delta_{3/2}}+\rme^{2\rmi\delta_{1/2}}-3)
 \cos\theta,
 \nonumber\\
 h &= -{2\pi\over\epsilon_{\bm k}}{1\over 2\rmi k^3}
 (-\rme^{2\rmi\delta_{3/2}}+\rme^{2\rmi\delta_{1/2}}).
\end{align}
Substituting these results into Eq.~\eqref{eq:spin-relaxation-Lambda} and performing the angular integration and the change of variable, we find that the spin relaxation rate of $\Lambda$ baryons is given by
\begin{align}
 \gamma_{\Lambda_s}
 &= {4\over 3\pi} \int_{m_{\pi}}^\infty \diff \omega 
 |\rme^{2\rmi\delta_{3/2}(\omega)}-\rme^{2\rmi\delta_{1/2}(\omega)}|^2 
 \nonumber \\
 &\qquad \times
 n_B(\omega) \big[1+n_B(\omega)\big],
 \label{eq:gamma-Lambda-spin}
% \nonumber \\
% \tilgamma_\Lambda
% &= {1\over 2\pi} \int_{m_{\pi}}^\infty \diff\omega
% \Big(
%  |2\rme^{2\rmi\delta_{3/2}(\omega)}+\rme^{2\rmi\delta_{1/2}(\omega)}-3|^2
% \nonumber \\
% &\hspace{30pt}
%  +2 |\rme^{2\rmi\delta_{3/2}(\omega)}-\rme^{2\rmi\delta_{1/2}(\omega)}|^2
% \Big) 
% % \nonumber \\
% % &\quad \times
% n_B(\omega) \big[1+n_B(\omega)\big], 
% \nonumber\\
% &= {3\over 2\pi} \int_{m_{\pi}}^\infty \diff \omega
% \left(
%  2|\rme^{2\rmi\delta_{3/2}(\omega)}-1|^2+|\rme^{2\rmi\delta_{1/2}(\omega)}-1|^2
% \right) 
% \nonumber \\
% &\quad \times
% n_B(\omega) \big[1+n_B(\omega)\big].
\end{align}
Thus, the remaining task is to find the phase shifts $\delta_{3/2}$ and $\delta_{1/2}$ for the $\Lambda \pi$ scattering.

As explained above, we assume that the phase shifts are dominated, either by a bound state near the threshold in the $J=1/2$ channel, or a resonance in the $J=3/2$ channel, respectively. 
For the $J=3/2$ channel, we use the standard Breit-Wigner form of the $S$-matrix corresponding to the resonance $\Sigma^*(1385)$:
\be
 S_{3/2}\equiv \rme^{2\rmi\delta_{3/2}(\omega)}
 = {\omega-\omega_0-\rmi\Gamma/2\over \omega-\omega_0+\rmi\Gamma/2}.
\ee
We will use the mass diference $\omega_0=m_{\Sigma^*}-m_\Lambda \approx 269$ MeV and the decay width $\Gamma=36$ MeV for the known $\Sigma^*(1385)$ resonance found in, e.g., the PDG data book~\cite{ParticleDataGroup:2020ssz}.
This enables us to set the phase shift $\delta_{3/2}$.

For the $J=1/2$ channel, we employ the simplest effective-range approximation for the P-wave $S$-matrix as
\be
 S_{1/2} \equiv 
 \rme^{2\rmi\delta_{1/2}(k)}
 \simeq
 {\rmi k-{1\over a^3 k^2}-{1\over r}\over -\rmi k-{1\over a^3 k^2}-{1\over r}},
 \label{eq:S-1/2}
\ee
with the two length parameters, $a$ and $r$, which denote the P-wave scattering length and the effective range, respectively. 
This form is consistent with the analyticity requirement of the $S$-matrix in the complex $k$-plane, i.e., $S(-k)=S(k)^{-1}$ and $S(-k^*)=S(k)^*$, as well as the low $k$ behavior of the phase shift in the $l=1$ channel, i.e., $\delta(k)\sim k^{2l+1}=k^3$.

There is an important constraint on $(a,r)$, which follows from the hadron spectroscopy, i.e., the $S$-matrix should have a single bound state pole at $k=\rmi\kappa$, representing the $\Sigma$ baryon.
Here, $\kappa=\sqrt{m_\pi^2-(m_\Sigma-m_\Lambda)^2}\approx 119$ MeV is the location of the $\Sigma$ baryon pole on the imaginary axis of the complex pion momentum $k=\sqrt{\omega^2-m_\pi^2}$. 
This bound state constraint takes the form
\be
\kappa+{1\over a^3\kappa^2}-{1\over r}=0, 
\label{eq:constraint}
\ee
which gives one relation between the two parameters $a$ and $r$.
Besides, a further constraint exists for $r$; a positive value of $r$ would not work since the above bound state equation for $\kappa$ then turns out to have an additional deeper bound state than the $\Sigma$ baryon, which does not exist in Nature. 
To proceed further, we choose three exemplar values for the effective range $r$, that is, $r=-1.0$ fm, $-0.5$ fm, and $-0.1$ fm. 
Solving the constraint \eqref{eq:constraint}, we find the corresponding scattering length determined as $a=-1.2$ fm, $-1.0$ fm, and $-0.64$ fm, respectively. 
Then, Eq.~\eqref{eq:S-1/2} gives us the phase shift $\delta_{1/2}$.

\begin{figure}[t]
 \includegraphics[width=0.95\linewidth]{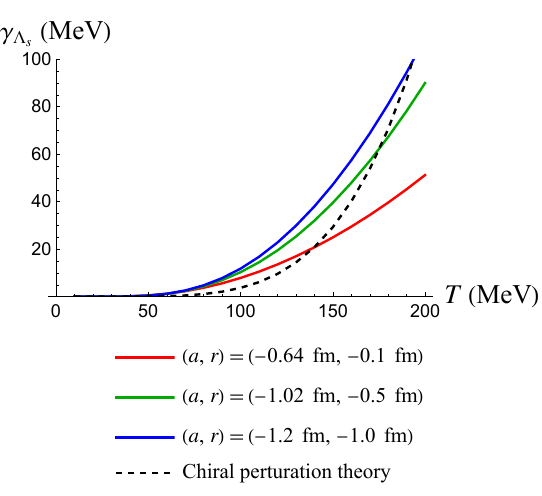}
 \caption{The spin relaxation rate of $\Lambda$ baryon as a function of temperature.
 Red, green, and blue solid lines are the results from the effective range theory with a choice of the printed parameters. 
 The Black dashed line is the result of the chiral perturbation theory with $\Lambda$ and $\Sigma$ baryons.
 }
 \label{fig7}
\end{figure}

With this phenomenological model for the scattering phase shifts, the spin relaxation rate of $\Lambda$ baryon as a function of temperature is plotted in Fig.~\ref{fig7}, for three different choices of the parameters $(a,r)$.
We note that the dependence of the $\Lambda$-baryon spin relaxation rate on different choices of parameters is not very significant numerically, although we observe the tendency of a larger spin relaxation rate for a larger value of $r$.
We find that the spin relaxation rate of the $\Lambda$ baryon around $T=150$ MeV is between $30$ MeV and $50$ MeV, corresponding to a relaxation time $\tau \simeq 4\sim7$ fm/c.

\section{Discussion\label{sec:Discussion}}

We find that the spin relaxation rate of baryons in the hadronic phase around temperature $T=150$ MeV is between $30$ MeV and $50$ MeV, corresponding to a relaxation time $\tau \simeq 4\sim7$ fm/c.
This time scale is comparable to the time scale of the plasma evolution in heavy-ion collisions, which indicates that an initial out-of-equilibrium spin polarization in the hadronic phase will remain so for a significant fraction of the time spent by the plasma before freezeout. 
This finding clearly demonstrates the necessity of including relaxation dynamics of spin polarization in any reliable theoretical prediction or phenomenological simulation for the spin polarization of observed hadrons.

In our analysis, we have used the properties of baryons and their interaction with pions as in the vacuum, without considering possible thermal corrections at finite temperature. 
For example, the mass and width of baryon resonances in a finite temperature plasma may significantly be different from their value in the vacuum. 
While for mesons such corrections have been studied in some detail previously \cite{Sommerer:1994tw}, we are not aware of a similar study for the baryon resonances in the temperature range we are interested in. 
This, among many other possible improvements in our treatment, can be a subject of future study.

\acknowledgements
We thank James Vary for helpful discussions on the properties of hadrons in finite temperature plasma.
We also thank Tetsuo Hatsuda and Tetsuo Hyodo for useful discussions on the hadron scattering.
This work is supported by the U.S. Department of Energy, Office of Science, Office of Nuclear Physics, grant No. DEFG0201ER41195 (MS, H-UY), and by Japan Society for the Promotion of Science (JSPS) KAKENHI Grant Nos.\ 21H01084 (YH) and 22K20369 (MH).
This work was partially supported by the RIKEN iTHEMS Program.

\appendix
\begin{widetext}

\section{Kadanoff-Baym approach to deriving kinetic equation}
\label{sec:Kadanoff-Baym}
	
In this appendix, as an alternative to Sec.~\ref{sec:spin-kinetic-theory}, we derive the quantum kinetic equation for the single baryon density matrix relying on the Kadanoff-Baym formalism~\cite{Kadanoff-Baym1961}.
In the Kadanoff-Baym formalism, we derive the equation of motion for the Wigner-transformed real-time Green function $S_{ab}$ (Keldysh indices $a=1,2$) of the baryon as
\begin{equation}
 \begin{split}
 \left( 
 \partial_t + \frac{1}{M} \bp \cdot \bnab_x
 \right)
 S_{12} 
 + [ \rmi \re \Sigma_{ra}, S_{12} ]_{\star}
 - [\Sigma_{12}, \im S_{ra}]_{\star}
 = - \frac{\rmi}{2}
 \Big[ 
 \{ \Sigma_{21}, S_{12} \}_{\star}
 - \{ \Sigma_{12}, S_{21} \}_{\star}
 \Big],
 \end{split}
 \label{eq:Boltzmann}
\end{equation}
where $\Sigma$ denotes the self-energy of baryons with the appropriate indices $1$ and $2$ ($r$ and $a$) on the Schwinger-Keldysh contour, $[A,B]_{\star}$ and $\{A,B\}_{\star}$  are the commutator and anti-commutator with the Moyal product.
Note that we use nonrelativistic Green functions for baryons with a $2\times2$ spin matrix, not a $4\times4$ Dirac matrix.

The relation between quantities with $12$ and $ra$ indices is, e.g., given by 
$S_{ra} = (S_{11} - S_{12} + S_{21} - S_{22})/2$ 
for the Green function (see, e.g., Appendix A of Ref.~\cite{Hongo:2022izs} for further details).
A reasonable approximation for the self-energy enables us to obtain the closed equation of motion for the Green function.
We note that the self-energy appearing in the right-hand side of Eq.~\eqref{eq:Boltzmann} includes collision processes.

Before computing the self-energy in our setup, let us first simplify the Kadanoff-Baym equation~\eqref{eq:Boltzmann}.
The collisions between the baryon and the background thermal pions is the dominant processes causing the baryon spin relaxation.
To capture these contributions at leading order, we perform the following successive approximations.
First of all, we simplify Eq.~\eqref{eq:Boltzmann} by neglecting the commutator terms in the left-hand side, since they represent higher order corrections for the energy dispersion relations, and also by replacing the Moyal products in the right-hand side with the usual product, in leading order of the gradient expansion.
We also neglect the spatial dependence of $S_{12}$ in leading gradient expansion, which results in the equation
\begin{equation}
 \partial_t S_{12} (t,p)
 = - \frac{1}{2}
 \Big[ 
 \{ \rmi \Sigma_{21} (t,p), S_{12} (t,p) \}
 - \{ \rmi \Sigma_{12} (t,p), S_{21} (t,p) \}
 \Big].
\label{eq:eom-S12}
\end{equation}
We employ the standard quasi-particle approximation for the baryon Green's functions as 
\begin{equation}
 \begin{split}
 S_{12} (t,p) 
 &= - \rho (t,\bp) 2\pi \delta (p^0-E_{\bp}),
 \\
 S_{21} (t,p) 
 &= \big[\bm{1}_{2\times 2} - \rho (t,\bp)\big] 
 2\pi \delta (p^0-E_{\bp}),
 \end{split}
 \label{eq:S12-rho}
\end{equation}
where $\rho (t,\bp)$ is identified as the $2\times 2$ spin density matrix as introduced in Eq.~\eqref{eq:rho}, and $E_{\bp}=\bp^2/(2M_N)$ is the energy of the baryon with mass $m_N$.

We now specify the relevant leading order diagrams for the self-energy.
The vital point here is that the two-body baryon-pion elastic scatterings are the dominant processes in a low-temperature pion gas~\cite{Leutwyler:1990uq}.
We then identify the self-energy diagrams shown in Fig.~\ref{fig-self-energy}, which correctly captures the dominant scattering processes.
The solid and dashed lines denote the baryon and the pion Green's functions, respectively, while the blob represents a vertex function,
$t^{ba}_{\beta\alpha}(p',k';p,k)$, which reduces to the $\matrixT$-matrix for a process $B^\alpha(\bm p)+\pi^a(\bm k)\to B^\beta(\bm p')+\pi^b(\bm k')$, when we require the on-shell codition for all particles (the indices $\alpha,\beta=p,n,\Lambda$ specify the baryon species, and $a,b=1,2,3$ the pion isospin triplet states).
Assuming that the baryon is nonrelativistic, we have a relation between $t^{ba}_{\beta\alpha}$ and the $\matrixT^{ba}_{\beta\alpha}$ in the main text (recall Eq.~\eqref{eq:rela-nonrela}) as 
\begin{equation}
 t^{ba}_{\beta\alpha} (p',k';p,k)
 \big|_{\mathrm{on~shell}}
 = 2 \epsilon_{\bk} 
 \matrixT^{ba}_{\beta\alpha} (\bm p',\bm k';\bm p,\bm k),
 \label{eq:relation-t-T}
\end{equation}
with $\epsilon_{\bk} = \sqrt{\bk^2+m_\pi^2}$.
It should be emphasized that $t^{ba}_{\beta\alpha}$ (and $T^{ba}_{\beta\alpha}(\bm p',\bm k';\bm p,\bm k)$) is a $2 \times 2$ matrix in the spin space.
We do not include a modification of the pion propagator, $D_{ab}(k)$, due to higher-order interactions, and we have
\begin{equation}
 \begin{split}
 D_{12} (k) 
 &= n_B (k^0) \frac{2 \pi }{2 \epsilon_{\bk}}
 \left[
 \delta (k^0 - \epsilon_{\bk})
 - \delta (k^0 + \epsilon_{\bk})
 \right], 
 \\
 D_{21} (k) 
 &= \big[1+n_B (k^0)\big] 
 \frac{2 \pi }{2 \epsilon_{\bk}}
 \left[
 \delta (k^0 - \epsilon_{\bk})
 - \delta (k^0 + \epsilon_{\bk})
 \right],
 \end{split}
 \label{eq:propagators-pion}
\end{equation}
where $n_B(k^0)=1/(\exp( k^0/T)+1)$ is the Bose-Einstein distribution function with the temperature $T$.
 
The explicit expression for the self-energy following from the Feynman rules is given by
\begin{equation}
 \begin{split}
 - \rmi \Sigma_{12}^{\beta} (p)
 &= \frac{1}{2} \int_{k,k',p'} \sum_{a,b,\alpha}
 t_{\beta\alpha}^{ba} (p,k;p',k')
 S_{12}^{\alpha} (p')
 [t_{\beta\alpha}^{ba} (p,k;p',k')]^\dag
 D_{21}^{a} (k) D_{12}^{b} (k')
 (2\pi)^4 \delta^{(4)} (p+k-p'-k'),
 \\
 - \rmi \Sigma_{21}^{\beta} (p)
 &= \frac{1}{2} \int_{k,k',p'} \sum_{a,b,\alpha}
 t_{\beta\alpha}^{ba} (p,k;p',k')
 S_{21}^{\alpha} (p')
 [t_{\beta\alpha}^{ba} (p,k;p',k')]^\dag
 D_{12}^{a} (k) D_{21}^{b} (k')
 (2\pi)^4 \delta^{(4)} (p+k-p'-k')
 \end{split}
\end{equation}
% \begin{equation}
%  \begin{split}
%   - \rmi \Sigma_{12}^{pp} (p)
%   &= \int_{k,k',p'} \sum_{a,b}
%   T_{pp}^{ba} (\bp,\bk;\bp',\bk')
%   S_{12}^{pp} (p')
%   [T_{pp}^{ba} (\bp,\bk;\bp',\bk')]^*
%   D_{21} (k) D_{12} (k')
%   (2\pi)^4 \delta^{(4)} (p+k-p'-k')
%   \\
%   &\quad 
%   + \int_{k,k',p'} 
%   T_{np}^{ba} (\bp,\bk;\bp',\bk')
%   S_{12}^{nn} (p')
%   [T_{np}^{ba} (\bp,\bk;\bp',\bk')]^*
%   D_{21} (k) D_{12} (k')
%   (2\pi)^4 \delta^{(4)} (p+k-p'-k'),
%   \\
%   - \rmi \Sigma_{21}^{pp} (p)
%   &= \int_{k,k',p'} \sum_{a,b}
%   T_{pp}^{ba} (\bp,\bk;\bp',\bk')
%   S_{21}^{pp} (p')
%   [T_{pp}^{ba} (\bp,\bk;\bp',\bk')]^*
%   D_{21} (k) D_{21} (k')
%   (2\pi)^4 \delta^{(4)} (p+k-p'-k')
%   \\
%   &\quad 
%   + \int_{k,k',p'} 
%   T_{np}^{ba} (\bp,\bk;\bp',\bk')
%   S_{12}^{nn} (p')
%   [T_{np}^{ba} (\bp,\bk;\bp',\bk')]^*
%   D_{21} (k) D_{12} (k')
%   (2\pi)^4 \delta^{(4)} (p+k-p'-k').
%  \end{split}
% \end{equation}
where we used a shorthand notation
$\int_{p_1, \cdots p_n}\equiv 
\int \frac{\diff^4 p_1}{(2\pi)^4} \cdots \frac{\diff^4 p_n}{(2\pi)^4}$.
Substituting these into Eq.~\eqref{eq:eom-S12}, neglecting $O(\rho^2)$-terms for the dilute baryon system, and performing $p^0$-integration using the energy $\delta$-function, we eventually obtain the following equation,
\begin{equation}
 \begin{split}
 \partial_t \rho^{\beta} (t,\bp)
 &= \frac{1}{2}
 \int \frac{\diff p^0}{2\pi}
 \Big[ 
 \{-\rmi \Sigma_{12}^\beta (t,p), S_{21}^\beta (t,p) \}
 - \{- \rmi \Sigma_{21}^\beta (t,p), S_{12}^\beta (t,p) \}
 \Big]
 \\
 &= \frac{1}{2} \frac{1}{2}
\int \frac{\diff p^0}{2\pi}
 \int_{k,k',p'} \sum_{a,b,\alpha}
 \Big[ 
 \big\{
 t_{\beta\alpha}^{ba} (p,k;p',k')
 S_{12}^{\alpha} (p')
 [t_{\beta\alpha}^{ba} (p,k;p',k')]^\dag,
 S_{21}^\beta (t,p)
 \big\}
 \\
 &\hspace{70pt} \times
 D_{21}^{a} (k) D_{12}^{b} (k')
 (2\pi)^4 \delta^{(4)} (p+k-p'-k')
 \\
 &\hspace{72pt}
 - \big\{
 t_{\beta\alpha}^{ba} (p,k;p',k')
 S_{21}^{\alpha} (p')
 [t_{\beta\alpha}^{ba} (p,k;p',k')]^\dag,
 S_{12}^\beta (t,p)
 \big\}
 \\
 &\hspace{70pt} \times
 D_{12}^{a} (k) D_{21}^{b} (k')
 (2\pi)^4 \delta^{(4)} (p+k-p'-k')
 \Big]
 \\
 &\simeq 
 - \frac{1}{2}
 \int_{\bk,\bk',\bp'} \sum_{a,b,\alpha}
 \Big[ 
 \big\{
 \matrixT_{\beta\alpha}^{ba} (\bp,\bk;\bp',\bk')
 [-\rho^\alpha (t,\bp') ] 
 [\matrixT_{\beta\alpha}^{ba} (\bp,\bk;\bp',\bk')]^\dag,
 \bm{1}_{2\times2}
 \big\}
 \\
 &\hspace{70pt} \times
 n_B (\epsilon_{\bk'}) \big[ 1 + n_B (\epsilon_{\bk}) \big]
 (2\pi)^4 \delta^{(3)} (\bp+\bk-\bp'-\bk')
 \delta (E_{\bp}+\epsilon_{\bk}-E_{\bp'}-\epsilon_{\bk'})
 \\
 &\hspace{72pt}
 - \big\{
 \matrixT_{\beta\alpha}^{ba} (\bp,\bk;\bp',\bk')
 \bm{1}_{2\times2} [\matrixT_{\beta\alpha}^{ba} (\bp,\bk;\bp',\bk')]^*,
 [- \rho^\beta (t,\bp)]
 \big\}
 \\
 &\hspace{70pt} \times
 n_B (\epsilon_{\bk}) \big[ 1 + n_B (\epsilon_{\bk'}) \big]
 (2\pi)^4 \delta^{(3)} (\bp+\bk-\bp'-\bk')
 \delta (E_{\bp}+\epsilon_{\bk}-E_{\bp'}-\epsilon_{\bk'})
 \Big]
 \\
 &= \int_{\bk,\bk',\bp'} \sum_{a,b,\alpha}
 \Big[ 
 \matrixT_{\beta\alpha}^{ba} (\bp,\bk;\bp',\bk')
 \rho^\alpha (t,\bp') 
 [\matrixT_{\beta\alpha}^{ba} (\bp,\bk;\bp',\bk')]^\dag
 \\
 &\hspace{70pt} \times
 n_B (\epsilon_{\bk'}) \big[ 1 + n_B (\epsilon_{\bk}) \big]
 (2\pi)^4 \delta^{(3)} (\bp+\bk-\bp'-\bk')
 \delta (E_{\bp}+\epsilon_{\bk}-E_{\bp'}-\epsilon_{\bk'})
 \\
 &\hspace{72pt}
 - |\matrixT_{\beta\alpha}^{ba} (\bp,\bk;\bp',\bk')|^2
 \rho^\beta (t,\bp)
 \\
 &\hspace{70pt} \times
 n_B (\epsilon_{\bk}) \big[ 1 + n_B (\epsilon_{\bk'}) \big]
 (2\pi)^4 \delta^{(3)} (\bp+\bk-\bp'-\bk')
 \delta (E_{\bp}+\epsilon_{\bk}-E_{\bp'}-\epsilon_{\bk'})
 \Big],
 \end{split}
\end{equation}
where we used Eqs.~\eqref{eq:relation-t-T}-\eqref{eq:propagators-pion} to obtain the third line, and also employed the center-of-mass frame in which $\epsilon_{\bk} = \epsilon_{\bk'}$ is satified.
To obtain the last line, we noticed that
$|\matrixT_{\beta\alpha}^{ba} (\bp,\bk;\bp',\bk')|^2$ after $\bk$ and $\bk'$ integrations commutes with $\rho$, because it is proportional to the identity matrix thanks to rotational symmetry.
With appropriate identification of the baryon indices, these equations are identical to the kinetic equations~\eqref{time} and \eqref{lambdatime} in the main text.
	
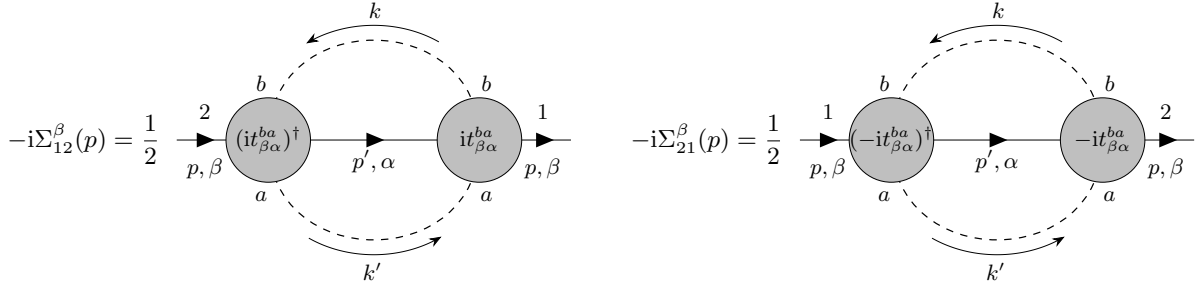
\begin{figure}
 \centering
\includegraphics[width=0.9\linewidth]{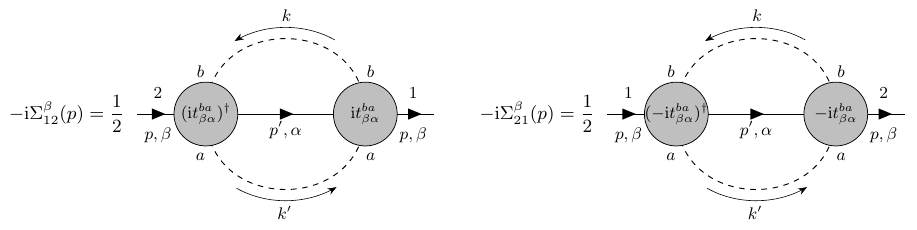}  
\cout{
 $- \rmi \Sigma_{12}^{\beta} (p)= \dfrac{1}{2}$
 \scalebox{0.9}{
 \begin{tikzpicture}[baseline=(o.base)]
 \begin{feynhand}
 \vertex (o) at (0,-0.1) {};
 \vertex [dot] (f1) at (1.5,0);
 \vertex [dot] (f2) at (-1.5,0);
 \vertex (f10) at (2.8,0);
 \vertex (f20) at (-2.8,0);
 \node at (2.4,0.4) {$1$};
 \node at (-2.4,0.4) {$2$};
 \node at (2.4,-0.4) {$p,\beta$};
 \node at (-2.4,-0.4) {$p,\beta$};
 \node at (1.6,0.8) {$b$};
 \node at (-1.6,0.8) {$b$};
 \node at (1.6,-0.8) {$a$};
 \node at (-1.6,-0.8) {$a$};
 \node at (0,-0.3) {$p',\alpha$};
 \propag [with arrow=0.7] (f1) to (f10);
 \propag [with arrow=0.3] (f20) to (f2);
 \propag [with arrow=0.5] (f2) to (f1);
 \propag [sca,mom'={[arrow shorten=0.3] $k$}] (f1) to [out=90,in=90, looseness=1.618] (f2);
 \propag [sca,revmom={[arrow shorten=0.3] $k'$}] (f1) to [out=270,in=270, looseness=1.618] (f2);
 \filldraw[fill=lightgray] (1.5,0) circle [radius=6mm];
 \filldraw[fill=lightgray] (-1.5,0) circle [radius=6mm];
 \vertex at (-1.5,0) {{\small $(\rmi t_{\beta\alpha}^{ba})^\dag$}};
 \vertex at (1.5,0) {{\small $\rmi t_{\beta\alpha}^{ba}$}};
 \end{feynhand}
 \end{tikzpicture}}
 \qquad 
 $- \rmi \Sigma_{21}^{\beta} (p)= \dfrac{1}{2}$
 \scalebox{0.9}{
 \begin{tikzpicture}[baseline=(o.base)]
 \begin{feynhand}
 \vertex (o) at (0,-0.1) {};
 \vertex [dot] (f1) at (1.5,0);
 \vertex [dot] (f2) at (-1.5,0);
 \vertex (f10) at (2.8,0);
 \vertex (f20) at (-2.8,0);
 \node at (2.4,0.4) {$2$};
 \node at (-2.4,0.4) {$1$};
 \node at (2.4,-0.4) {$p,\beta$};
 \node at (-2.4,-0.4) {$p,\beta$};
 \node at (1.6,0.8) {$b$};
 \node at (-1.6,0.8) {$b$};
 \node at (1.6,-0.8) {$a$};
 \node at (-1.6,-0.8) {$a$};
 \node at (0,-0.3) {$p',\alpha$};
 \propag [with arrow=0.7] (f1) to (f10);
 \propag [with arrow=0.3] (f20) to (f2);
 \propag [with arrow=0.5] (f2) to (f1);
 \propag [sca,mom'={[arrow shorten=0.3] $k$}] (f1) to [out=90,in=90, looseness=1.618] (f2);
 \propag [sca,revmom={[arrow shorten=0.3] $k'$}] (f1) to [out=270,in=270, looseness=1.618] (f2);
 \filldraw[fill=lightgray] (1.5,0) circle [radius=6mm];
 \filldraw[fill=lightgray] (-1.5,0) circle [radius=6mm];
 \vertex at (-1.5,0) {{\small $(-\rmi t_{\beta\alpha}^{ba})^\dag$}};
 \vertex at (1.5,0) {{\small $-\rmi t_{\beta\alpha}^{ba}$}};
 \end{feynhand}
 \end{tikzpicture}}
 }
 \caption{The self-energy diagrams that capture the two-body baryon-pion scattering.}
 \label{fig-self-energy}
\end{figure}

\section{Chiral perturbation theory for \texorpdfstring{$\Lambda\pi$}{Lambda-pi} scattering}
\label{sec:chiPT}

At sufficiently low temperature, the baryon-pion scattering is reliably described by the chiral perturbation theory.
Here we provide a derivation of the low-energy expression of $(g,h)$ in Eq.~\eqref{eq:T-matrix-lambda-pi} with the parameters in the chiral perturbation theory~\cite{Weinberg:1978kz,Gasser:1983yg,Gasser:1984gg}.

First of all, we note that the pion is an isospin triplet pseudo scalar meson with negative parity while the $\Lambda$- and $\Sigma$-baryons are isospin singlet and triplet $J^P = 1/2^+$ baryons with one strange quark.
Taking into account that pions are pseudo Nambu-Goldstone bosons, we identify the leading-order chiral Lagrangian relevant for the $\Lambda\pi$ scattering as (see, e.g, Ref.~\cite{Scherer-Schindler2011} for details)
\begin{equation}
 \Lcal_{\Lambda\pi}
 = \frac{1}{2} (\partial_\mu \pi^a)^2
 - \frac{1}{2} m_\pi^2 (\pi^a)^2
 + \pSig^a (\rmi \gamma^\mu \partial_\mu - m_\Sigma) \Sigma^a
 + \pLam (\rmi \gamma^\mu \partial_\mu - m_\Lambda) \Lambda
 + \frac{D}{\sqrt{3}{f_\pi}} 
 \big[ 
  \pSig^a \gamma^\mu \gamma_5 \Lambda \partial_\mu \pi^a 
  + {\rm h.c.} \big],
 \label{eq:chiral-L-LO}
\end{equation}
where $\pi^a,\Sigma^a~(a=1,2,3)$ and $\Lambda$ denote the pion, $\Sigma$-baryon, and $\Lambda$-baryon fields with the corresponding masses $m_\pi = 140\,$MeV, $m_\Sigma=1190\,$MeV and $m_\Lambda =1116\,$MeV, respectively.
$\gamma^\mu$, and $\gamma_5$ are the gamma matrices satisfying $\gamma^\mu \gamma^\nu+\gamma^\nu\gamma^\mu=2\mathop{\mathrm{diag}}(1,-1,-1,-1)$ and $\gamma_5=\rmi\gamma^0\gamma^1\gamma^2\gamma^3$.
We also introduced the pion decay constant $f_\pi = 90\,$MeV and a dimensionless coupling constant $D$, which one can determine as $D=0.80$ from a semileptonic decay process~\cite{Borasoy:1998pe}.

From Eq.~\eqref{eq:chiral-L-LO}, we identify learding-order processes in the $\Lambda\pi$ scattering, diagramatically shown in Fig.~\ref{fig-Lambda-pi-scattering}.
Computing these tree graphs, we find the relativistic scattering amplitude given by
\begin{align}
  \rmi \Mcal_{\Lambda \pi}
 &= \rmi \delta_{ab} \frac{4 \bar{m} D^2}{3f_\pi^2}
  \pu (p') 
  \left[
  1
  + \bar{m} \Delta m 
  \left( 
   \frac{1}{s - m_\Sigma^2} + \frac{1}{u - m_\Sigma^2}
  \right)
  - \frac{\bar{m}}{2} 
  \gamma^\mu(k_\mu + k'_\mu) 
  \left( 
   \frac{1}{s - m_\Sigma^2}
  - \frac{1}{u - m_\Sigma^2}
  \right)
  \right] u(p)
  \nonumber \\
  &= \rmi \delta_{ab} \frac{4 \bar{m} D^2}{3f_\pi^2}
   \pu (p') 
   \bigg[
   1
   + \frac{\bar{m} \Delta m }{s - m_\Sigma^2} 
   + \frac{\bar{m} \Delta m }{u - m_\Sigma^2}
   + \left(
   \frac{s-u}{4m_\Lambda} 
   - \rmi \sigma^{\mu\nu} \frac{k_\mu k'_\nu}{2m_\Lambda}
   \right)
   \left( 
   - \frac{\bar{m}}{s - m_\Sigma^2}
   + \frac{\bar{m}}{u - m_\Sigma^2}
   \right)
   \bigg] 
   u(p),
\end{align}
where we introduced $\Delta m \equiv m_\Sigma - m_\Lambda$, $2\bar{m} \equiv m_\Sigma + m_\Lambda$, $\sigma^{\mu\nu}=\rmi(\gamma^\mu\gamma^\nu-\gamma^\nu\gamma^\mu)/2$,
Mandelstam variables 
$s \equiv (p^0+k^0)^2 - (\bp+\bk)^2$ and $u \equiv (p^0-k'^0)^2 - (\bp-\bk')^2$, 
and the wave function for the $\Lambda$-baryon $u(k)$ ($\bar{u}(k)=u^\dag(k)\gamma^0$).
To derive the second line, we used the Gordon decomposition 
$\bar{u} (p') \gamma^\mu u(p) 
 = \bar{u} (p') 
 \left[
  {p^\mu + p'^\mu} 
  + \rmi \sigma^{\mu\nu} (p'_\nu-p_\nu)
 \right] u(p)/(2m_N)$
as well as the energy-momentum conservation $p+k=p'+k'$.
Note that we parametrize the $S$-matrix element as 
$\langle \bm{p}',\bm{k}'|(S-1)|\bm{p},\bm{k}\rangle =  (2\pi)^4 \delta^{(4)} (p+k-p'-k') \rmi\Mcal_{\Lambda\pi}$.

To proceed further, let us take the center-of-mass frame, in which 
$\bk+\bp=\bm{0}=\bk'+\bp'$ and $\epsilon_{\bk} + E_{\bk} = \epsilon_{\bk'} + E_{\bk'}$, and thus, $|\bk| = |\bk'|$ is satisified.
Note that two Mandelstam variables are simplified to be
$s = (E_{\bk} + \epsilon_{\bk})^2$ and 
$u = (E_{\bk} - \epsilon_{\bk})^2 - 2 \bk^2 (1+\cos \theta)$
in the center-of-mass frame. 
Moreover, considering the heavy-baryon limit, we approximate the $\Lambda$-baryon wave function as
$u(p) \simeq ( \sqrt{2m_\Lambda} \chi, 0)^t$ with the two-component (or nonrelativistic) spinor $\chi^{(s)}$, which results in
\begin{align}
 \rmi \Mcal_{\Lambda \pi}
  &\simeq \rmi \delta_{ab} \frac{8 m_\Lambda \bar{m} D^2}{3f_\pi^2}
   \chi^\dag
   \bigg[
   1
   + \frac{\bar{m} \Delta m }{s - m_\Sigma^2} 
   + \frac{\bar{m} \Delta m }{u - m_\Sigma^2}
   - \frac{\bar{m}}{4m_\Lambda} 
    \left( 
    \frac{s-u}{s - m_\Sigma^2}
    - \frac{s-u}{u - m_\Sigma^2}
    \right)
  \nonumber \\
   &\hspace{85pt}
   - \rmi \bm{\sigma} \cdot (\bk' \times \bk)
   \frac{\bar{m}}{2m_\Lambda}
   \left( 
    \frac{1}{s - m_\Sigma^2} - \frac{1}{u - m_\Sigma^2}
   \right)
   \bigg] 
   \chi.
\end{align}
Then, recalling Eq.~\eqref{eq:rela-nonrela}, we find the $\matrixT$-matrix for the $\Lambda\pi$ scattering predicted from the chiral perturbation as follows:
\begin{align}
 \matrixT_{\Lambda \pi} 
 &\simeq
  - \delta_{ab} \frac{2 \bar{m} D^2}{3f_\pi^2 \epsilon_{\bk}}
   \bigg[
   1
   + \frac{\bar{m} \Delta m }{s - m_\Sigma^2} 
   + \frac{\bar{m} \Delta m }{u - m_\Sigma^2}
   - \frac{\bar{m}}{4m_\Lambda} 
    \left( 
    \frac{s-u}{s - m_\Sigma^2}
    - \frac{s-u}{u - m_\Sigma^2}
    \right)
  \nonumber \\  &\hspace{64pt}
   - \rmi \bm{\sigma} \cdot (\bk' \times \bk)
   \frac{\bar{m}}{2m_\Lambda}
   \left( 
   \frac{1}{s - m_\Sigma^2}
   - \frac{1}{u - m_\Sigma^2}
   \right)
   \bigg] .
\end{align}
By matching this result with Eq.~\eqref{eq:T-matrix-lambda-pi}, we find the low-momentum expansion of the $\matrixT$-matrix~\eqref{eq:g-h-chiPT} in the main text.

\begin{figure}
 \centering
\includegraphics[width=0.6\linewidth]{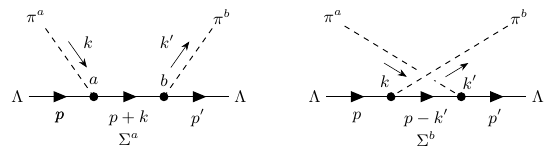}  
\cout{
 \scalebox{0.9}{
    \begin{tikzpicture}[baseline=(o.base)]
     \begin{feynhand}
      \vertex (o) at (0,-0.1) {};
      \vertex (f1) at (2,0) {};
      \vertex (f2) at (-2,0) {};
      \vertex (v1) [dot] at (0.66,0) {};
      \vertex (v2) [dot] at (-0.66,0) {};
      \vertex (d1) at (1.65,1.4) {};
      \vertex (d2) at (-1.65,1.4) {};
      \node at (-1.75,1.5) {$\pi^a$};
      \node at (1.75,1.5) {$\pi^b$};
      \node at (-1.3,-0.4) {$p$};
      \node at (1.3,-0.4) {$p^{\prime}$};
      \node at (-2.1,0) {$$};
      \node at (-2.1,0) {$\Lambda$};
      \node at (2.1,0) {$\Lambda$};
      \node at (0,-0.8) {$\Sigma^a$};
      \node at (-1.3,-0.4) {$p$};
      \node at (0,-0.4) {$p+k$};
      \node at (0.66,0.3) {$b$};
      \node at (-0.66,0.3) {$a$};
      \propag [fer] (v1) to (f1);
      \propag [fer] (v2) to (v1);
      \propag [fer] (f2) to (v2);
      \propag [sca, revmom'={[arrow shorten=0.3] $k'$}] (d1) to (v1);
      \propag [sca, mom={[arrow shorten=0.3] $k$}] (d2) to (v2);
     \end{feynhand}
    \end{tikzpicture}} 
 \qquad 
  \scalebox{0.9}{
     \begin{tikzpicture}[baseline=(o.base)]
      \begin{feynhand}
       \vertex (o) at (0,-0.1) {};
       \vertex (f1) at (2,0) {};
       \vertex (f2) at (-2,0) {};
       \vertex (v1) [dot] at (0.66,0) {};
       \vertex (v2) [dot] at (-0.66,0) {};
       \vertex (d1) at (1.65,1.4) {};
       \vertex (d2) at (-1.65,1.4) {};
       \node at (-1.75,1.5) {$\pi^a$};
       \node at (1.75,1.5) {$\pi^b$};
       \node at (-2.1,0) {$\Lambda$};
       \node at (2.1,0) {$\Lambda$};
       \node at (0,-0.8) {$\Sigma^b$};
       \node at (-1.3,-0.4) {$p$};
       \node at (1.3,-0.4) {$p^{\prime}$};
       \node at (0,-0.4) {$p -k^{\prime}$};
       \propag [sca, mom'={[arrow shorten=0.4] $k$}] (d2) to (v1);
       \propag [sca, revmom={[arrow shorten=0.4] $k'$}, top] (d1) to (v2);
       \propag [fer] (v1) to (f1);
       \propag [fer] (v2) to (v1);
       \propag [fer] (f2) to (v2);
      \end{feynhand}
     \end{tikzpicture}}
}     
 \caption{The leading-order contribution to the $\Lambda\pi$ scatetering in the chiral perturbation theory.} 
 \label{fig-Lambda-pi-scattering}
\end{figure}
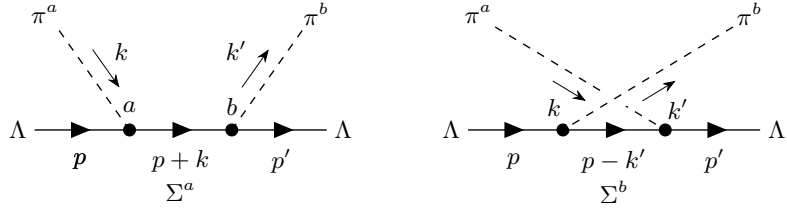

\end{widetext}

\bibliography{spin-damping}

\end{document}